\documentclass[pdflatex,sn-mathphys-num]{sn-jnl}

\usepackage{graphicx}%
\usepackage{multirow}%
\usepackage{amsmath,amssymb,amsfonts}%
\usepackage{amsthm}%
\usepackage{mathrsfs}%
\usepackage[title]{appendix}%
\usepackage{xcolor}%
\usepackage{textcomp}%
\usepackage{manyfoot}%
\usepackage{booktabs}%
\usepackage{float}
\usepackage{algorithm}%
\usepackage{algorithmicx}%
\usepackage{algpseudocode}%
\usepackage{listings}%
\usepackage{subfigure}
\usepackage{subcaption}
\usepackage{tikz}
\usetikzlibrary{shapes.geometric, arrows}

\tikzstyle{startstop} = [rectangle, rounded corners, minimum width=3cm, minimum height=1cm, text centered, draw=black, fill=red!30]
\tikzstyle{process} = [rectangle, minimum width=3cm, minimum height=1cm, text centered, draw=black, fill=orange!30]
\tikzstyle{decision} = [diamond, minimum width=3cm, minimum height=1cm, text centered, draw=black, fill=blue!30]
\tikzstyle{arrow} = [thick,->,>=stealth]

\theoremstyle{thmstyleone}%
\theoremstyle{thmstyletwo}%

\theoremstyle{thmstylethree}%

\begin{document}

\title[Investigation of cloud cavitating flow in a venturi using Adaptive Mesh Refinement (AMR)]{Investigation of cloud cavitating flow in a venturi using Adaptive Mesh Refinement (AMR)}


\author*[1]{\fnm{Dhruv} \sur{Apte}}\email{dhruvga@vt.edu}

\author[2]{\fnm{Mingming} \sur{Ge}}\email{mmge@vt.edu}

\author[1,3]{\fnm{Olivier} \sur{Coutier-Delgosha}}\email{ocoutier@vt.edu}

\affil*[1]{\orgdiv{Kevin T. Crofton Department of Aerospace and Ocean Engineering}, \orgname{Virginia Tech},
\orgaddress{\city{Blacksburg}, \postcode{24060}, \state{VA}, \country{USA}}}

\affil[2]{\orgdiv{Macao Environmental Research Institute, Faculty of Innovation Engineering}, \orgname{Macau University of Science and Technology}, \orgaddress{\city{Macao}, \postcode{999078}, \state{Macao SAR}, \country{China}}}

\affil[3]{\orgname{Univ. Lille, CNRS, ONERA, Arts et Metiers ParisTech, Centrale Lille}, \orgaddress{ \city{Lille}, \postcode{F-59000}, \country{France}}}


\abstract{Unsteady cloud cavitating flow is detrimental to the efficiency of hydraulic machinery like pumps and propellers due to the resulting side-effects of vibration, noise and erosion damage. Modelling such a unsteady and highly turbulent flow remains a challenging issue. In this paper, cloud cavitating flow in a venturi is calculated using the Detached Eddy Simulation (DES) model combined with the Merkle model. The Adaptive Mesh Refinement (AMR) method is employed to speed up the calculation and investigate the mechanisms for vortex development in the venturi. The results indicate the velocity gradients and the generalized fluid element strongly influence the formation of vortices throughout a cavitation cycle. In addition, the cavitation-turbulence coupling is investigated on the local scale by comparing with high-fidelity experimental data and using profile stations. While the AMR calculation is able to predict well the time-averaged velocities and turbulence-related aspects near the throat, it displays discrepancies further downstream owing to a coarser grid refinement downstream and under-performs compared to a traditional grid simulation . Additionally, the AMR calculations is unable to reproduce the cavity width as observed in the experiments. Therefore, while AMR promises to speed the process significantly by refining grid only in regions of interest, it is comparatively in line with a traditional calculation for cavitating flows. Thus, this study intends to provide a reference to employing AMR as a tool to speed up calculations and be able to simulate turbulence-cavitation interactions accurately.}

\keywords{cavitating flow, detached eddy simulation (DES), cavitation model, adaptive mesh refinement}



\maketitle

\section{Introduction}\label{sec1}

Cavitation is a multiphase, highly unstable and turbulent phenomenon characterized by formation of clouds of bubbles when the ambient pressure drops below the vapour pressure. These clouds of bubbles travel with high velocity and upon exiting the low-pressure area burst, generating shock. This shock can cause erosion damage, vibration, load asymmetry and other detrimental effects that can drastically impact the performance of marine engineering devices like propellers, pumps and turbines. Additionally, the capability to cause erosion can also be utilized in jet drilling in the hydrocarbon and geothermal energy sector \cite{chi2022erosion, wu2023structure}. Therefore, it is important to investigate the phenomenon, especially the mechanism of periodic development of the main cavity followed by its shedding and collapse.

A primary component of cavitation is its interaction with turbulence and the resulting generated vortices. Gopalan \& Katz \cite{gopalan2000flow} used PIV and high-speed photography to demonstrate that vortice generation was a result of the collapse of the vapor cavities in the cavity closure region. More recently, Arabnejad \textit{et al.} \cite{arabnejad2019numerical} investigated the cavitation over the leading edge of a NACA0009 hydrofoil using High-Speed Visualization (HSV) to observe the transformation of the shedding of cavity structures into horse-show vortices. Ge \textit{et al.} \cite{ge2022combined} conducted similar experiments in a venturi-type cavitation reactor and observed a similar  horseshoe structure. They further concluded that the horseshoe broke up into a long, thin chain of bubbles with the upper parts merging with the clouds downstream. 

Concurrently, studies using numerical simulations to model cavitating flows have been gaining more and more interest. A successful modelling of cavitating flows typically necessitates the combination of a cavitation model and a turbulence model. Regarding turbulence models, Direct Numerical Simulation (DNS) where all turbulence scales are resolved seems an evident approach but its demand for an extremely fine mesh and computational power hinders its application in practice. Large Eddy Simulation (LES) studies seem as a viable option and several studies have been conducted to investigate the cavity and vortex dynamics driving cavitating flows \cite{chen2017large, li2016large, liu2023investigation}. In fact, LES has also been coupled with multi-scale Euler-Lagrangian approaches \cite{wang2021euler} to investigate both the large vapor volumes from an Eulerian point-of-view and the micro-scale bubble dynamics using a Lagrangian reference. It was observed that the collision of the re-entrant jet with the cavity interface results in the formation of large number of bubbles. Tian \textit{et al.} \cite{tian2022multiscale} further confirmed these findings with their multi-scale approach and concluded that the generation of the large number of bubbles is a result of Kelvin-Helmholtz instability, triggered by the shear flow between water and vapor. Ji \textit{et al.} \cite{ji2024cavitation} have conducted a comprehensive review of multi-scale techniques for simulating cavitating flows. However, LES also requires high-resolution grids which results in higher computational costs \cite{luo2012numerical,li2021multiscale}. Thus, Reynolds-averaged Navier-Stokes (RANS) models, where all turbulence scales are modelled seems as the practical option to model cavitating flows. 

Coutier-Delgosha \textit{et al.} \cite{coutier2003evaluation} conducted RANS calculations using the k-$\epsilon$ Re-Normalization Group (RNG) and k-$\omega$ models with and without an empirical correction suggested by Reboud \textit{et al.} \cite{reboud1998two} to simulate cavitating flows. They concluded that the standard models over-predict the eddy-viscosity and thus are unable to predict the periodic shedding of the cavity as observed in experiments. The same model was also employed by Ji \textit{et al.} \cite{ji2014numerical} to simulate the unsteady cavitating flow around a twisted hydrofoil. They noted that the cavitating flow induced boundary layer separation and reported large increases of vorticity at the cavity interface. Long \textit{et al.} \cite{long2017numerical} advanced this approach by combining the Reboud correction with the Filter-Based Method (FBM) \cite{johansen2004filter}, a recent type of model, termed as hybrid RANS-LES models that behave as RANS model near the wall and as LES model away from the wall, therefore combining the accuracy of LES models and computational efficiency of RANS models. They conducted a vorticity budget analysis, based on the Vorticity Transport Equation  (VTE) in attached cavitation around a Clark-Y hydrofoil. They concluded that the baroclinic torque term in the VTE is the principal source of vorticity production during the collapse of the cavitation cloud. Sun \textit{et al.} \cite{sun2019numerical} conducted a similar calculation using Partially-Averaged Navier-Stokes model (PANS) proposed by Girimaji \textit{et al.} \cite{girimaji2006partially} to simulate unsteady cavitation around a NACA0015 hydrofoil in thermo-sensitive fluid. They observed that the vortex structures and motion show considerable unsteady properties in the cavity shedding region at the trailing edge. Additionally, Apte \textit{et al.} \cite{apte2023numerical} conducted a systematic analysis of several RANS and hybrid RANS-LES models to simulate cloud cavitating flow inside a venturi nozzle and concluded that while these models are able to simulate the unsteady cavity shedding, they are unable to simulate the cavity dynamics on the local scale. They suggested grid refinement as a possible solution to ensure the LES zone occurs on the fine grid. 

To solve the problem of the resulting high requirement in computational resources, studies have proposed mesh refinement techniques that promote refinement in specific flow regions rather than refining the entire domain. Bai \textit{et al.} \cite{bai2018large} devised a Cartesian cut-cell method where the rotation factor, the ratio of the strain rate to the sum of the strain rate and vorticity tensor was applied as a marker for cell refinement while modelling a tip leakage cavitating flow generated by a hydrofoil. The methodology was successfully extended to Cheng \textit{et al.} \cite{cheng2020large} who conducted a systematic analysis of various LES models to simulate tip-vortex cavitation. They concluded the method was able to simulate tip vortex cavitation accurately and were able to capture the influence of the vorticity dilatation term in decreasing the vorticity inside the tip-leakage vortex section.  Another technique, titled Adaptive Mesh Refinement (AMR) was proposed \cite{berger1989local} that allocates high resolution grid based on the input property and can be a promising technique to understand better the vortex dynamics in cavitating flows. AMR has been used for various multiphase flow applications like primary atomization \cite{fuster2009simulation}, bubble dynamics \cite{zhang2018coalescence} and cavitation as well. Li \textit{et al.} \cite{li2020large} used AMR to aid LES for simulating cavitation over a cylindrical body. Wang \textit{et al.} \cite{wang2020numerical}, employed a similar approach to model cavitating flow around a Clark-Y hydrofoil. However, both these papers have leveraged LES, which continues to be prohibitively expensive in some case. In addition, while these studies delve into the aspects of cavitation generation mechanisms, there remains a gap to investigate the ability of AMR to capture the cavitation-turbulence interplay.  

This study aims to address this gap by simulating cloud cavitation using AMR coupled with the Detached Eddy Simulation (DES) model, a hybrid RANS-LES model. Thus, the study attempts to strike a delicate balance between computational time and accuracy. 

\section{Mathematical formulation and numerical method}
\subsection{Adaptive Mesh Refinement}
\begin{figure}[htb]
\centering
\begin{tikzpicture}[node distance=2.5cm, every node/.style={align=center}]

\node (start) [decision] {Refinement Interval?};
\node (criterion) [process, right of=start, xshift=4cm] {Refinement Criterion};
\node (engine) [process, below of=criterion] {Refinement Engine};
\node (refiner) [process, below of=engine] {Mesh Refiner};
\node (tree) [process, below of=refiner] {Refinement Tree};
\node (solver) [process, left of=engine, xshift=-5cm, yshift=-2.5cm, fill=green!30] {Solver};

\draw [arrow] (start.east) -- node[anchor=south] {Yes} (criterion.west);
\draw [arrow] (criterion) -- (engine);
\draw [arrow] (engine) -- (refiner);
\draw [arrow] (refiner) -- (tree);
\draw [arrow] (tree.west) -| ++(-6cm,0) -- (solver.south);
\draw [arrow] (start.south) -- ++(0,-2cm) -| node[anchor=north, near start] {No} (solver.north);

\end{tikzpicture}
\caption{Workflow for AMR method}
\label{fig:workflow}
\end{figure}
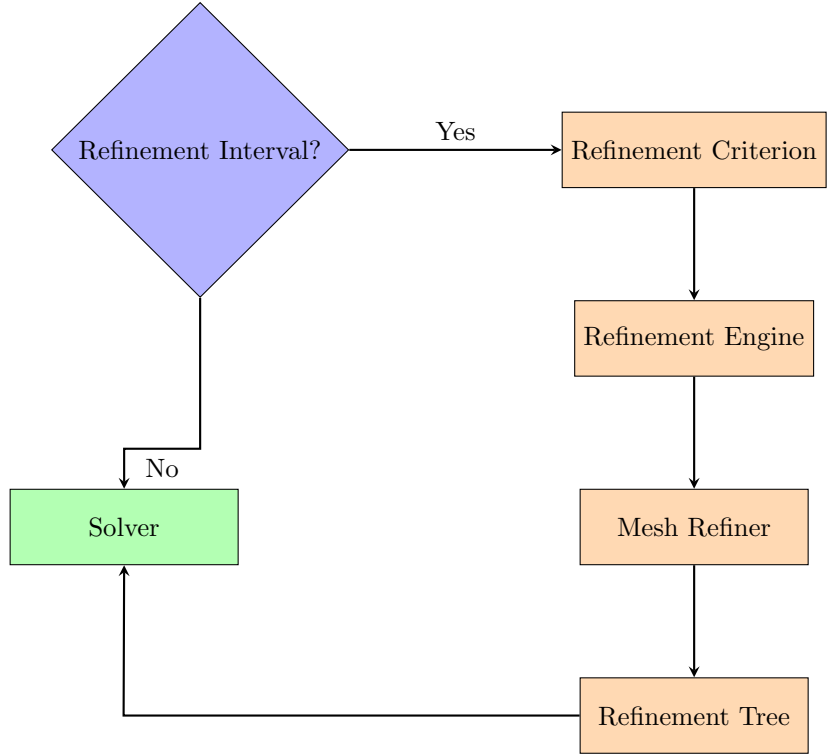

To overcome the issue of using highly-refined grids, the AMR is used to focus the refinement dynamically in the region most needed. Mathematically, AMR can be formulated as refining a domain comprising of non-overlapping rectangular grid $G_{l,k}$:
\begin{equation}
    G_{l} = \cup_{m} G_{l,k}
\end{equation}
where $l=0, 1, 2, ... l_{max}$ represents refinement levels. In OpenFOAM \cite{Weller1998}, AMR is based upon three key drivers, the refinement engine, mesh cutter and the refinement tree. The refinement engine is responsible for ensuring the connectivity structures in the mesh while the mesh cutter refines the mesh as per the input. Finally, the refinement tree saves the refinement history and ensures a steady transition for mesh refinement. Formulated by Mitchell \textit{et al.} \cite{mitchell2007refinement}, the refinement tree is defined as:
\begin{equation}
   T (G) = \{ \Omega, \{C(\nu_{i})\}\} 
\end{equation}
which contains a set of nodes $\Omega = {\nu_{i}}_{i=0}^{M}$ with each $\nu_{i} \in \Omega$ containing a set of child $C(\nu_{i}) \subset \Omega$ where G represents the grid. Fig \ref{fig:workflow} presents the workflow of the three key portions in AMR. The solver reads the refinement criterion as the solver reaches the refinement interval . This refinement criterion could be based on a numerical domain for turbulent kinetic energy or pressure field or as in this case, the void fraction. Next, the solver proceeds to the refinement engine which selects the appropriate zones for grid refinement. The mesh cutter then refines the specific areas while the refinement tree saves the entire refinement history. The refinement occurs in the form of an octree structure which means, on every single refinement level, the number of cells increases by a factor of eight. After refinement, a similar unrefinement process is conducted for cells that no longer fit the refinement criteria, with the solution being mapped to the new cells. The refinement criterion is based on the void fraction to ensure the grid refinement happens in the interface between the cavitating flow and single-phase, turbulent flow.  

\subsection{Numerical models}
\subsubsection{Basic Governing Equations}
As previously mentioned,modelling cavitating flows involves a coupling of a cavitation model and a turbulence model. The work here uses the Transport-Equation Model approach (TEM) where the two phases are considered to be strongly coupled and governed by the same momentum and mass transfer equations:
\begin{eqnarray}
\frac{\partial (\rho_{m} u_{i})}{\partial t}+\frac{\partial (\rho_{m} u_{i} u_{j})}{\partial x_{j}}\nonumber\\
=-\frac{\partial p}{\partial x_{i}}+\frac{\partial}{\partial x_{j}}\left((\mu_{m} +\mu_{t})\left(\frac{\partial u_{i}}{\partial x_{j}}+\frac{\partial u_{j}}{\partial x_{i}}- \frac{2}{3}\frac{\partial u_{k}}{\partial x_{k}} \delta_{ij}\right)\right)
\end{eqnarray}

\begin{equation}
    \frac{\partial\rho_{l} \alpha_{l}}{\partial t}+\frac{\partial(\rho_{l} \alpha_{l} u_{j})}{\partial x_{j}}= \dot{m}^{-} + \dot{m}^{+}
\end{equation}
\begin{equation}
    \rho_{m}= \rho_{l}\alpha_{l}+\rho_{v}\alpha_{v}
    \label{eq:mixture_density}
\end{equation}
\begin{equation}
    \mu_{m}= \mu_{l}\alpha_{l}+\mu_{v}\alpha_{v}
\end{equation}
where $u_{j}$ is the velocity component in the jth direction, $\rho_{m}$ and $\mu_{m}$ are respectively the density and viscosity of the mixture phase, \textit{u} is the velocity, \textit{p} is the pressure, $\rho_{l}$ and $\rho_{v}$ are respectively the liquid and vapor density, $\mu_{l}$ and $\mu_{v}$ are respectively the liquid and vapor dynamic viscosity while $\mu_{t}$ represents the turbulent viscosity. $\alpha_{l}$ and $\alpha_{v}$ are respectively the liquid and vapor void fraction. The source ($\dot{m}^{+}$) and sink ($\dot{m}^{-}$) terms represent the condensation (vapor destruction) and evaporation (vapor formation) terms respectively and will be discussed below.

The separation of condensation and evaporation terms in the above equation is the principal concept of the TEM methodology. The separation aids in treating the two processes separately if needed. The TEM model used in the study is the Merkle model \cite{merkle1998computational} which defines the evaporation and condensation terms as:
\begin{equation}
    \dot{m}^{-}= \frac{C_{dest} min(p-p_{sat},0)\gamma \rho_{l}}{0.5 U_{\infty}^{2} t_{\infty}\rho_{v}}
\end{equation}
\begin{equation}
    \dot{m}^{+}= \frac{C_{prod} max(p-p_{sat},0)(1-\gamma)}{0.5 U_{\infty}^{2} t_{\infty}}
\end{equation}
where $\gamma$ is the liquid volume fraction, $\rho_{v}$ and $\rho_{l}$ are the vapor density and the liquid density respectively, $p$ and $p_{sat}$ are the pressure and the saturation pressure respectively, $t_{\infty}$ is the free stream time scale and $U_{\infty}$ is the free stream velocity.  The empirical factors $C_{dest}$ and $C_{prod}$ are set as 80 and 1e-3 respectively, according to the study by Merkle \textit{et al.} \cite{merkle1998computational}.

\subsubsection{Turbulence model}

The turbulence model used is the Detached Eddy Simulation (DES) model with the k-$\omega$ SST model as the baseline RANS model \cite{travin2002physical}. Here, the LES model calculates the turbulent separated zone while the RANS model is used to model the near-wall zone. The switch to from RANS to LES and vice versa is defined by the change in length scale:
\begin{equation}
    L_{DES}= min (L_{t}, C_{DES} \Delta)
\end{equation}
where $L_{T}$ is the turbulent length scale defined as:
\begin{equation}
    L_{t} = \frac{\sqrt{k}}{\beta * \omega}
\end{equation}
and $C_{DES} \Delta$ is the DES filter length with $C_{DES}$ = 0.61

\section{Computational setup}

\begin{figure}[htb]
  \centering
\includegraphics[scale=0.3]{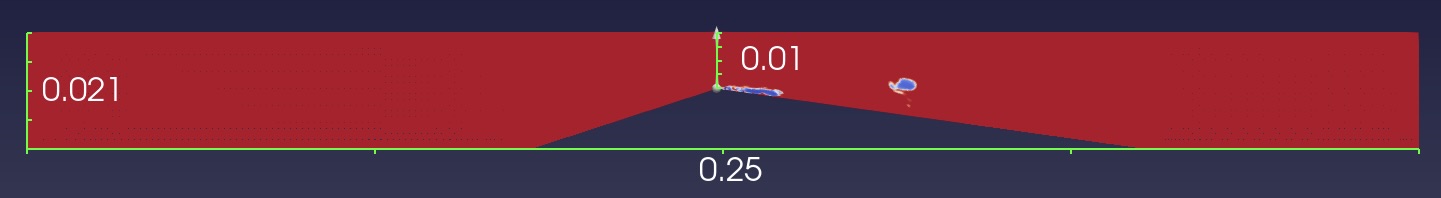}
  \caption{Schematic diagram of the computational geometry. The throat is where cavitation is initiated. All stated dimensions in metres. }
    \label{fig:geometry}
\end{figure}
\begin{figure}[htb]
  \centering
\includegraphics[scale=0.5]{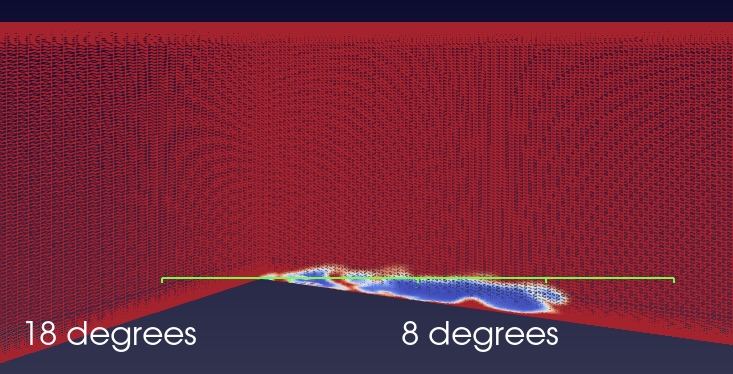}
  \caption{A zoomed-in diagram of the computational geometry highlighting the angles of the convergent-divergent section }
    \label{fig:geometry_angle}
\end{figure}

A converging-diverging (\textit{venturi}) nozzle is selected as the geometry utilized for the study as shown in \ref{fig:geometry}. The venturi nozzle has been widely used in jetting for geothermal reservoirs and hydrocarbon industry \cite{dehkordi2017cfd}. The geometrical setup here is identical to the experiment \cite{ge2021cavitation}.    The nozzle height at the inlet and outlet is originally 21 mm but steadily drops to 10 mm at the throat. The \textit{venturi} has a converging angle of 18 degrees and divergent angle of 8 degrees (see Fig \ref{fig:geometry_angle}). The velocity inlet and pressure outlet boundary conditions are adopted. For consistency and validation with the experiments, the velocity at inlet is set to 8.38 m/s while the pressure at the outlet is adjusted to ensure both experiments and simulations have the same mean cavity length. The Re number is 1.8 $\times 10^5$. To evaluate whether AMR is able to capture the cavity dynamics accurately at the local stage, a similar standard mesh calculation is conducted. The mesh consists of 8.4 million cells, since the cell limit of the AMR mesh is 8.4 million cells as well \cite{apte2024numerical}. Table \ref{tab:table1} shows the mean cavity lengths of the two cases, as compared to the experimental case. All three cases have the same mean cavity length, thus validating the simulations and providing an opportunity to investigate the cavitation-vortex interaction and the cavitation-turbulence interplay.

\begin{table}
    \centering
    \caption{\label{tab:table1} Mean cavity lengths for the two numerical cases as compared to experimental data}
    \begin{tabular}{lcc} \toprule
        Model & Cavity length (mm)\\ \midrule
         DES with AMR&  27\\
         DES 8.4 million (non-AMR) mesh& 27\\
         Experiment (\cite{ge2021cavitation})& 27 \\ \bottomrule
    \end{tabular}
\end{table}

\begin{figure}[!htb]
    \centering
    \begin{subfigure}
        \centering
        \includegraphics[width=\textwidth]{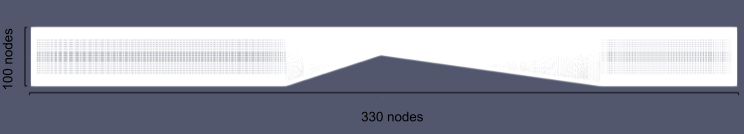}
    \end{subfigure}
    \hfill
    \begin{subfigure}
        \centering
        \includegraphics[width=0.8\textwidth]{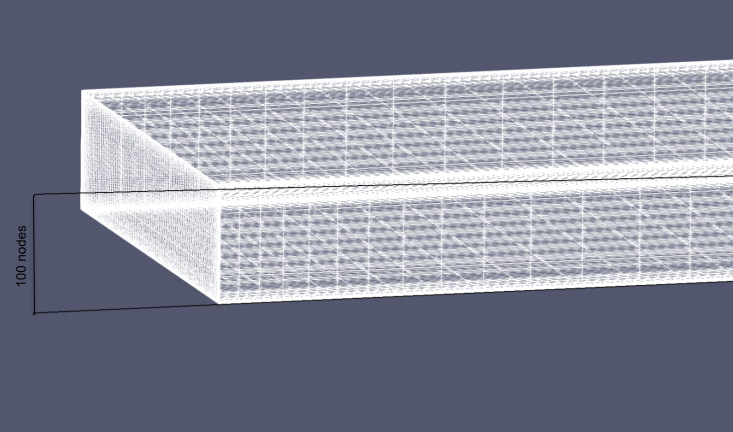}
    \end{subfigure}   
    \caption{Wireframe geometry highlighting the number of nodes or cells at the onset of the simulation. }
    \label{fig:domain_h}
\end{figure}

The calculations are conducted using \textit{interPhaseChangeFoam} and \textit{interPhaseChangeDyMFoam}.
\textit{interPhaseChangeFoam} is an unsteady, isothermal solver that couples a cavitation model with a turbulence model while the latter is its sister-solver that uses adaptive mesh refinement. It uses a \textit{dynamicMeshDict} for the dynamic mesh properties as input. The process of refining the grid is dependant on the field of void fraction, ranging from 0.01 to 0.9. This ensures the grids near the vapor cavity-water interface are refined. For this study, the mesh properties are based on the void fraction field with a maximum cell limit of 8.4 million cells from a base size of 3.3 million cells (see Fig \ref{fig:domain_h}). Both solvers utilize the PIMPLE algorithm, a hybrid algorithm comprising of both the PISO and SIMPLE algorithms used in OpenFOAM. The PIMPLE algorithm consists of three parts:
\begin{enumerate}
    \item momentum predictor
    \item pressure solver
    \item momentum corrector
\end{enumerate}
In PIMPLE, each timestep has at least one outer corrector loop where the fields are solved using a number of iterations. The algorithm exhibits a more robust pressure-velocity coupling by coupling a SIMPLE outer corrector loop with a PISO inner corrector loop and has been shown to display better numerical stability. 
Regarding numerical schemes, the volume fraction interpolation uses the Gauss vanLeer scheme. To mitigate the unboundedness, a semi-implicit  multi-dimensional limiter for explicit solution (MULES) is implemented: an implicit corrector step is first implemented corresponding to the discretization schemes and then an explicit correction is applied with the MULES limiter. The calculations are run at a fixed timestep of 1e-5. However, to ensure stability of the calculation, the AMR mesh is run at a timestep subjected to the maximum Courant number, 1 in this study. The \textit{refineInterval} utility is set to 10 thus refining the mesh at 10 timesteps. The refinement is based on the \textit{alpha.water} field or the void fraction field and set between the limits of 0.01 and 0.9. This ensures the mesh would be refined on lower void fraction values, where the cavity is expected to develop.

The simulation is set in three consecutive stages. First, it is run for 0.03 s where the vaporization coefficients in the cavitation model are set to zero thereby modelling it as a single-phase, turbulent flow. The next stage, consisting of 0.03s is a sinusoidal ramp whereupon at the end, fully cavitation has been fully launched. At the third and final stage, the fully cavitating regime is activated.  The focus of the work is on the fully cavitating regime with AMR also initiated at this stage. To maintain lower computational costs, the full cavitation regime is launched for 0.02 s, sufficient enough to capture a complete shedding cycle.

\section{Results}
\subsection{Cavitation flow structures}

\begin{figure}[htbp]
    \includegraphics[scale=0.25]{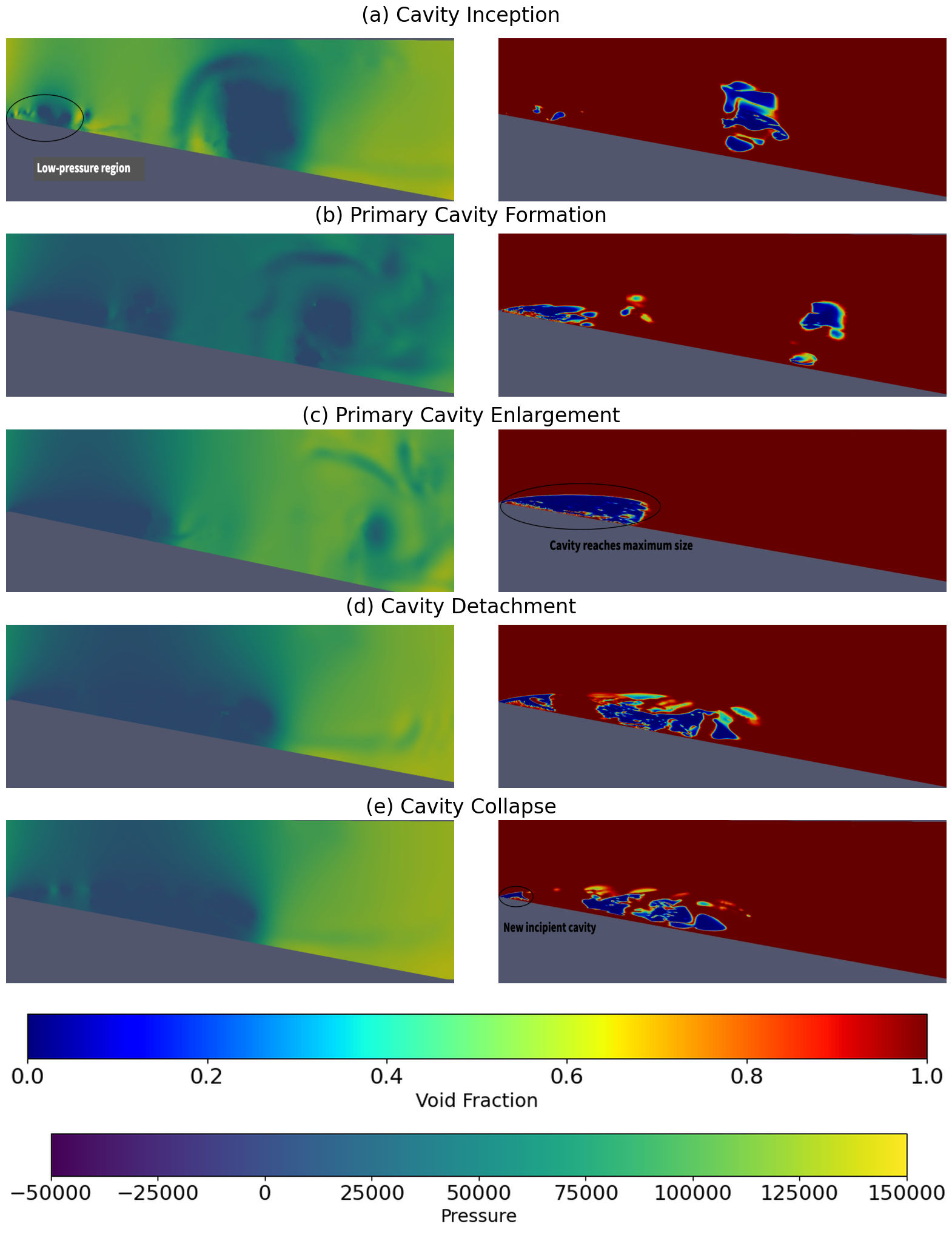}
    \caption{Pressure distributions and void fraction distribution snapshots for one cycle in the diverging section of the venturi. The first column shows the pressure distribution and the second column denotes the corresponding vapor distribution using void fraction.}
    \label{fig:p_grid}
\end{figure}
Fig \ref{fig:p_grid} shows the pressure and void fraction distributions at five successive time instants in the diverging section of the venturi. In the first column, the pressure distributions are shown, with the dark blue region denoting the low pressure regions. The second column represents the same snapshots with void fraction. Here, the blue region denotes vapor while the red region denotes water.  The distributions show the evolution of cavitation starting from the first one, where a cavity is initiated at the throat of the venturi. Near the throat, a sharp drop in pressure is observed, responsible for the inception of the cloud cavity. Downstream, a group of smaller cloud cavities is observed. These cavities are remnants of the previous cycle. The next figures Fig \ref{fig:p_grid} (c)-(d) demonstrate the growth of the cavity at the throat while the cavities downstream flow downstream and collapse as it leaves the low-pressure region. The next set of figures (e)-(f) show the cavity reaching its maximum size. In (g)-(h), it is observed that the large cavity at the throat is pinched-off due to a re-entrant jet rushing upstream, resulting in a secondary cloud cavity along with smaller vapor cavity clouds. The primary cavity continues to reduce and collapses while the secondary cavity continues to roll up and flow downstream. In (i)-(j), it is observed another small incipient cavity develops at the throat while the detached secondary cavity of this cycle continues to be shed downstream. Thus, the snapshots demonstrate the periodic nature of cloud cavitating flows and the presence of a re-entrant jet responsible for this periodic shedding of the primary cavity cloud.

\subsection{Adaptive Mesh Refinement inside the nozzle}

\begin{figure*}[h]
\centering
\includegraphics[width=\textwidth, height=6cm]{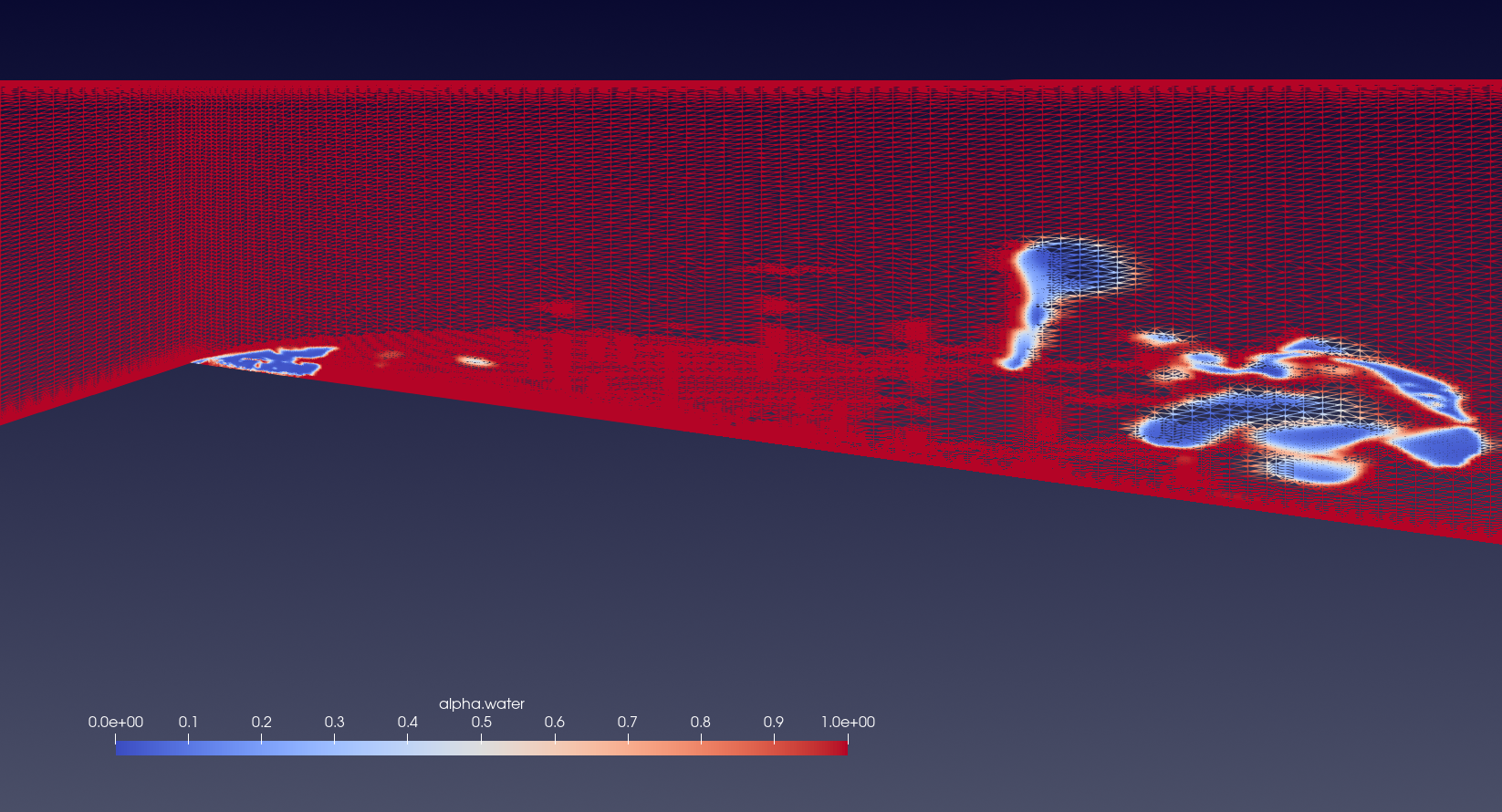}
\caption{Void fraction field at the cavity initiation stage in grid-mode. The blue indicates the vapor regions while the dominant red indicates water}
\label{fig:amr_ci}
\end{figure*}

\begin{figure*}[h]
\centering
\includegraphics[width=\textwidth, height=6cm]{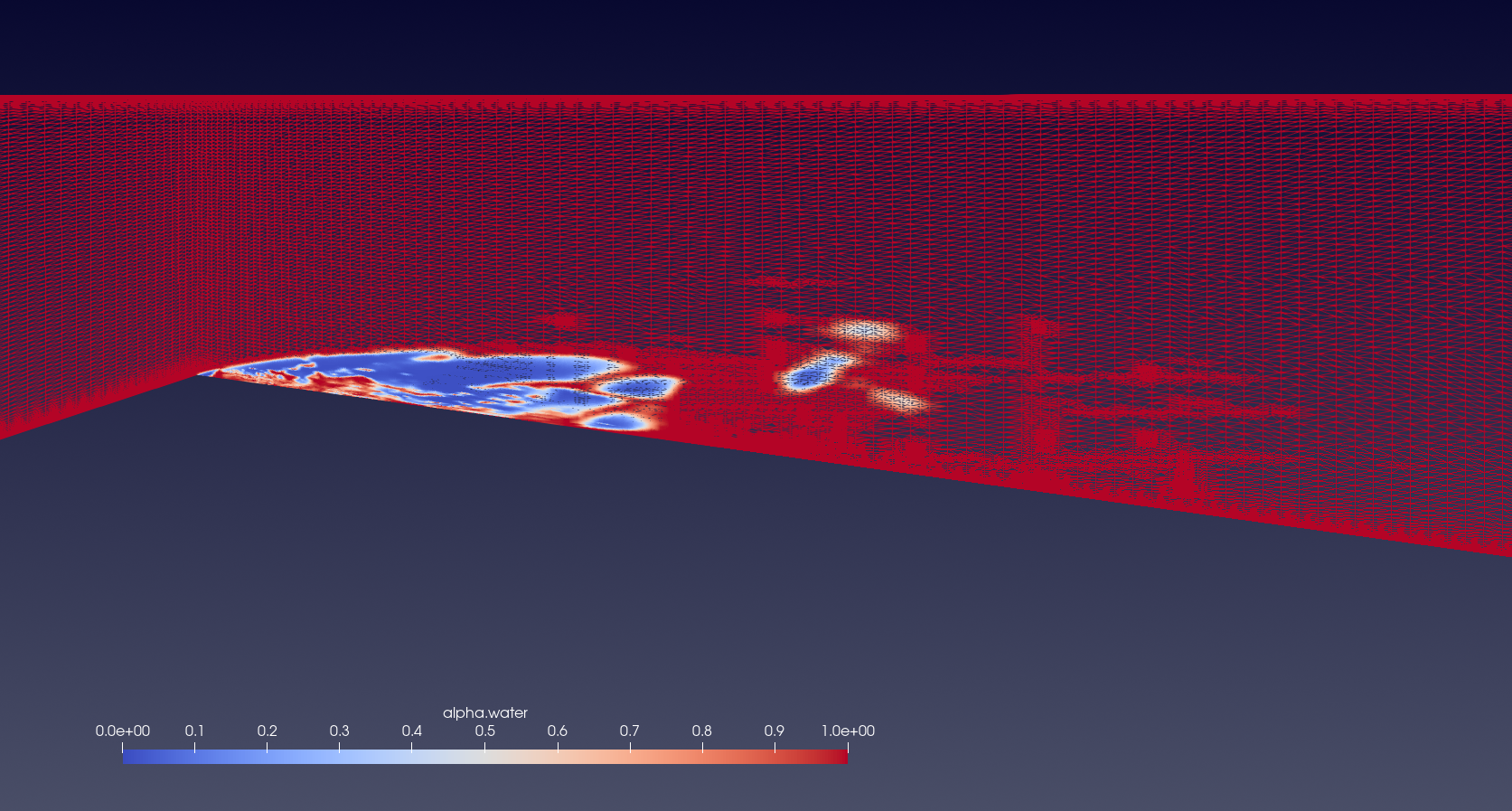}
\caption{Void fraction field at the cavity enlargement stage in grid-mode.The high refinement t the throat continues to grow due to the presence of a larger low-vapor region }
\label{fig:amr_ce}
\end{figure*}

\begin{figure*}[h]
\centering
\includegraphics[width=\textwidth, height=6cm]{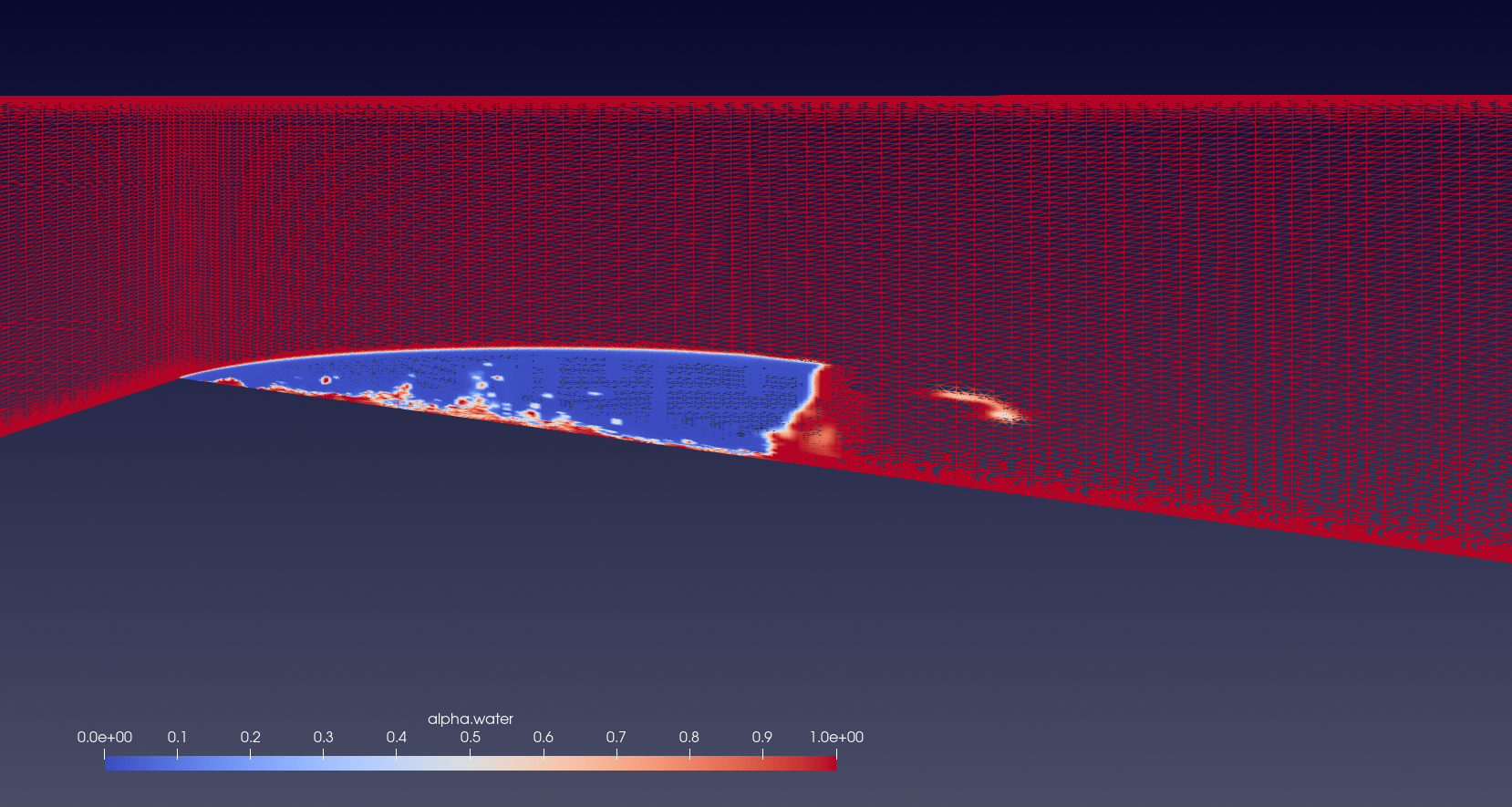}
\caption{Void fraction field as the cavity approaches its maximum length. It can be noted that the downstream region reverts back to its original structured grid due to lack of any vapor and thus, any refinement}
\label{fig:amr_cm}
\end{figure*}

As stated previously, the AMR method refines the grid only in the concerning region based on the input refinement criterion. To check the mesh refinement inside the venturi nozzle for unsteady cavitating flow, six snapshots are taken at the mid-plane of the venturi and focused on the diverging section of the nozzle. The grid is then visualized with the void fraction field. Fig \ref{fig:amr_ci} represents the grid during the cavity initiation stage. A small cavity at the throat is observed followed by a highly refined grid , in sharp contrast to the structured grid at the top of the venturi nozzle. The grid refinement extends further beyond the initial cavity, possibly as a result of other smaller cavities throughout the region. Further downstream, the refinement exists in selected regions, where the remnants of the previous cavity shedding cycle are flowing downstream. As the incipient cavity enlarges, shown in Fig \ref{fig:amr_ce}, the grid refinement closer to the throat is observed. However, further downstream where the detached cavities collapse after exiting the low-pressure region, the grid reverts back to the structured grid defined at the onset of the calculation. This hypothesis is substantiated in Fig \ref{fig:amr_cm} where the cavity reaches its maximum length. The grid in this region is significantly refined, while the grid downstream reverts completely to the structured grid devised as the original setup. Thus, it can be noted that the AMR method works well in refining solely the low vapor regions and giving more insights about the flow physics driving the unsteady cavitating flow at a much reduced computational cost.      

\begin{figure*}[htbp]
\centering
\includegraphics[width=\textwidth, height=6cm]{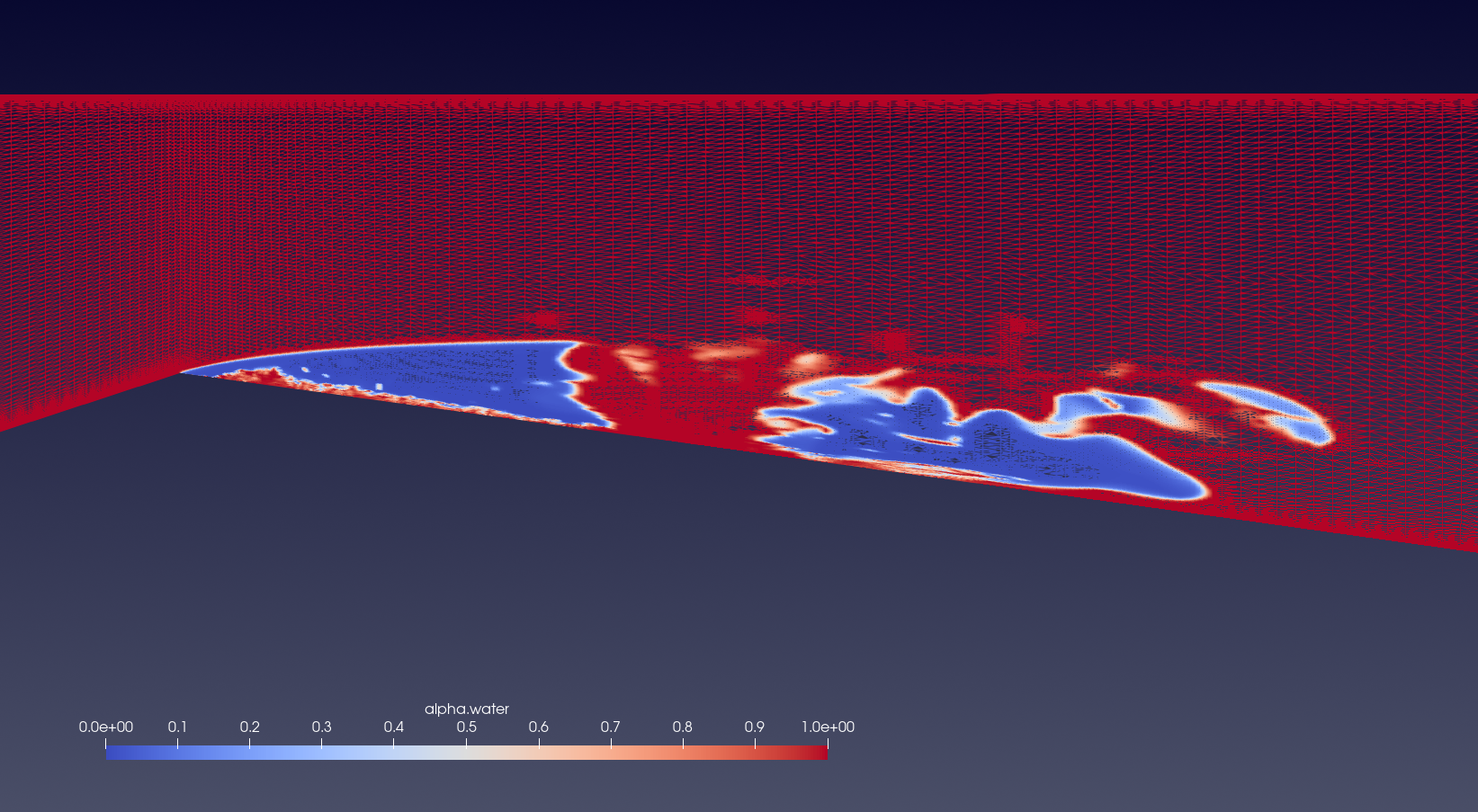}
\caption{Void fraction field as the primary cavity undergoes the detachment process. It is observed that a significant part of the mesh in the diverging section has now been refined, as a result of the primary cavity detaching into the secondary cloud cavity.}
\label{fig:amr_cd}
\end{figure*}
\begin{figure*}[htbp]
\centering
\includegraphics[width=\textwidth, height=6cm]{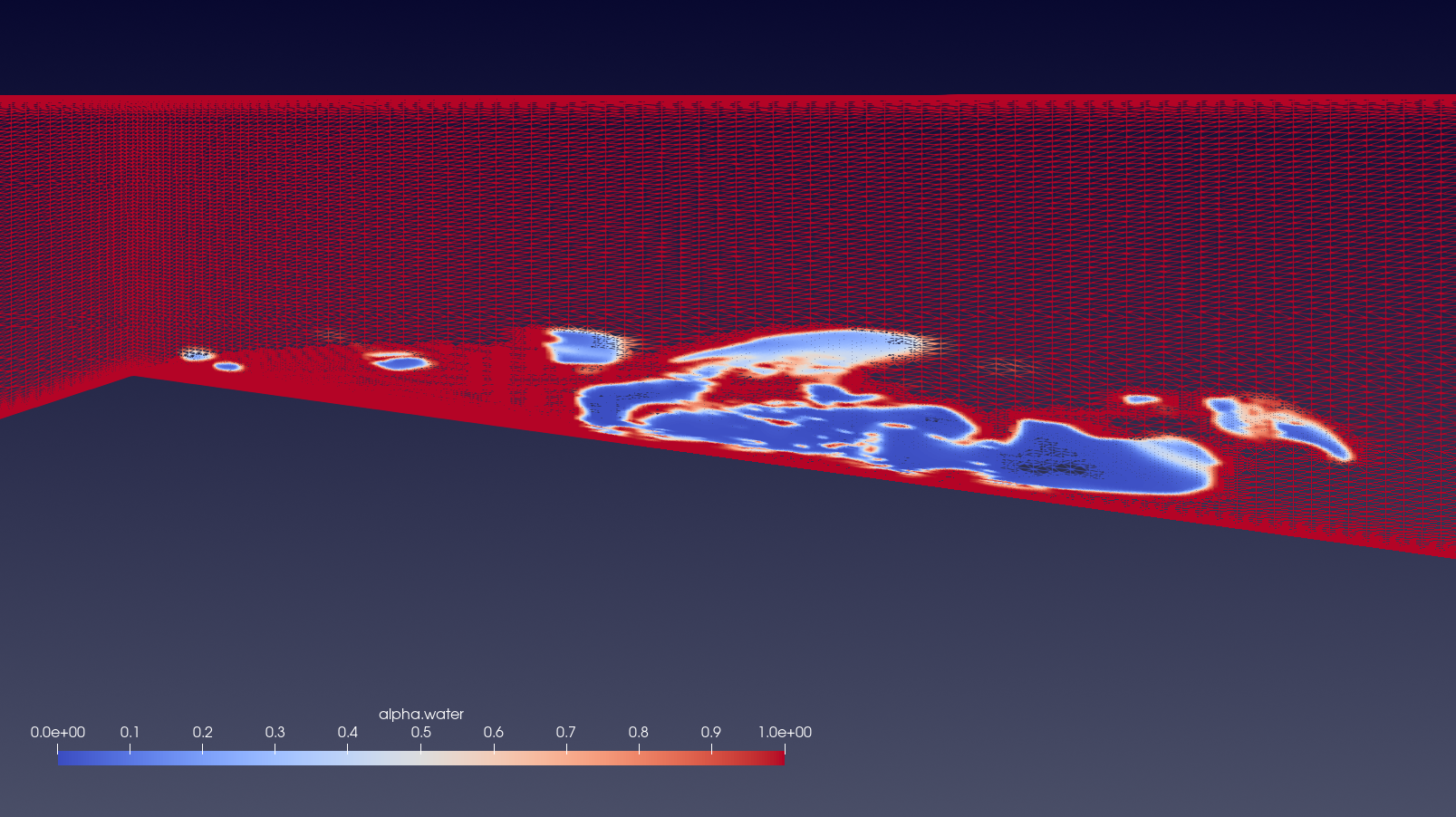}
\caption{Void fraction field as the primary cavity collapses with the downstream region continuing to be refined}
\label{fig:amr_cc}
\end{figure*}

Fig \ref{fig:amr_cd} shows the cavity detachment process. The presence of both a reduced primary cavity at the throat and a secondary cavity downstream results in a high grid refinement throughout the lower portion of the venturi nozzle, with some refinement zones slightly upwards as well, as a result of smaller shed vapor cavities in other planes. The cavity collapse process, depicted in Fig \ref{fig:amr_cc} shows the primary cavity collapsing and the detached cavity flowing downstream. Significant grid refinement continues to be conducted in the section accounted by the detached cavity downstream and the presence of smaller vapor cavities present closer to the throat. The presence of a refined grid in regions containing smaller vapor cavities illustrate that the AMR method is very efficient tool to simulate unsteady cavitating flows by dramatically increasing the grid refinement while striking a balance between grid refinement and computational cost.    

\subsection{Cavitation-vortex interaction}

For a better understanding of the cavitating flow and the resulting vortex interaction, the Q-criterion is plotted for various stages in a cavitating cycle. The Q-criterion defined as the second invariant of the velocity gradient tensor is defined as
\begin{equation}
    Q = \frac{1}{2} (|\Omega|^2 - |S|^2)
\end{equation}
where $\Omega$ is the vorticity tensor and \textit{S} is the rate of strain tensor. 
The Q-criterion and corresponding vapor structures are presented in the following figures. Here, a snapshot of the venturi from the front is taken of contours of the $\alpha$ = 0.6.
\begin{figure*}[h]
\centering
\includegraphics[width=\textwidth, height=10cm]{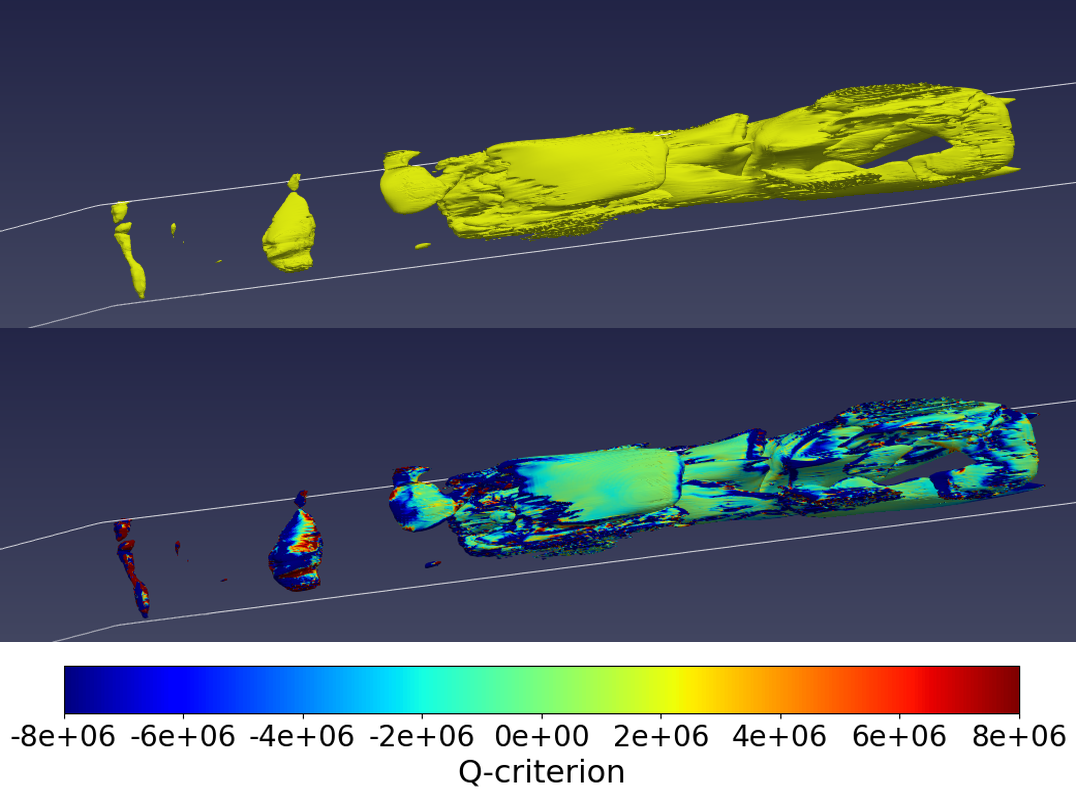}
\caption{Void fraction and Q-criterion contours for the cavity initiation stage}
\label{fig:ciq}
\end{figure*}

Fig \ref{fig:ciq} shows the void fraction and Q-criterion during cavitation inception. Here, the cavity is just initiated at the cavity throat while the remnants of the previous cavity appear further downstream including, the entire detached cloud cavity of the previous cycle.  
To further understand the mechanism of cavitation-vortex interaction, the vorticity-transport equation is employed at the mid-plane in the z-direction to investigate the vorticity and the distribution of its contributing terms, shown in the following equation

\begin{equation}
    \frac{D\omega_{z}}{Dt} = [(\omega \cdot \nabla) V]_{z} - [\omega (\nabla \cdot V)]_{z} +  \left( \frac{\nabla \rho_{m} \times \nabla p}{\rho_{m}^{2}} \right)_{z} + [(\nu_{m} + \nu_{t}) \nabla^{2} \omega]_{z} 
\end{equation}\label{eq:VTE}

Here,the Left Hand Side (LHS) denotes the rate of vorticity change while the Right Hand Side (RHS) indicate the vortex stretching, vortex dilatation, baroclinic torque and viscous diffusion of vorticity terms respectively. The vortex stretching term describes the stretching and tilting of a vortex due to the velocity gradients. The vortex dilatation describes the expansion and contraction of a fluid element. The baroclinic torque is a result of the misalignment between the pressure and velocity gradients. The viscous diffusion term can be ignored in high Reynolds number flows \cite{dittakavi2010large} and thus is omitted in the study. The equation clearly shows the effects of velocity and pressure gradients formed as a result of cavitating flow on the vorticity. The equation is further simplified as:
\begin{equation}
    \omega_{z} = \frac{\partial V_{y}}{\partial x} - \frac{\partial V_{x}}{\partial y}
\end{equation}
\begin{equation}
    [(\omega \cdot \nabla) V]_{z} = \omega_{x} \frac{\partial V_{z}}{\partial x} + \omega_{y} \frac{\partial V_{z}}{\partial y} + \omega_{z} \frac{\partial V_{z}}{\partial z}
\end{equation}
\begin{equation}
    [\omega (\nabla \cdot V)]_{z} = \omega_{z} (\frac{\partial V_{x}}{\partial x}+\frac{\partial V_{y}}{\partial y}+ \frac{\partial V_{z}}{\partial z})
\end{equation}
\begin{equation}
     \left( \frac{\nabla \rho_{m} \times \nabla p}{\rho_{m}^{2}} \right)_{z} = \frac{1}{\rho_{m}^{2}} (\frac{\partial \rho_{m}}{\partial x} \cdot \frac{\partial p}{\partial y} - \frac{\partial \rho_{m}}{\partial y} \cdot \frac{\partial p}{\partial x})
\end{equation}

\begin{figure*}[h]
\centering
\includegraphics[width=\textwidth, height=10cm]{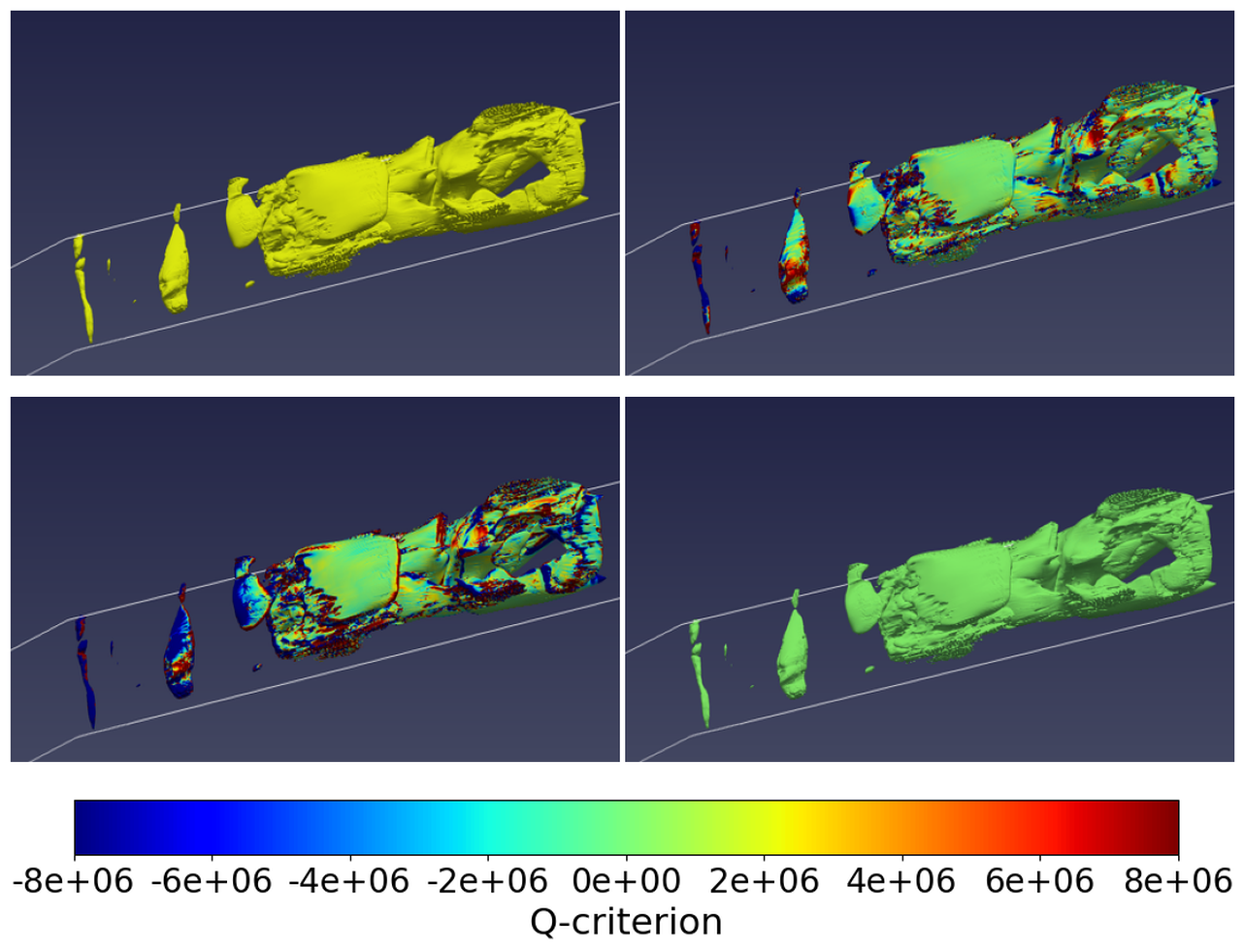}
\caption{Starting top left, in clockwise direction, contours of void fraction (contour of $\alpha$ = 0.6), vortex stretching, baroclinic torque, and vortex dilatation terms as an incipient cavity starts developing at the venturi throat.}
    \label{fig:ci}
\end{figure*}

Fig \ref{fig:ci} shows the different terms of the vorticity transport equation at the cavity initiation stage. It is observed, that while the vortex stretching and dilatation terms show highly positive values, specially at the throat, thus dominating the vortex formation. The baroclinic torque does not influence the vorticity considerably and it is demonstrated that the velocity gradients and expansion of the fluid element influences the cavitation-vortex interaction. 

\begin{figure*}[h]
\centering
\includegraphics[width=\textwidth, height=14cm]{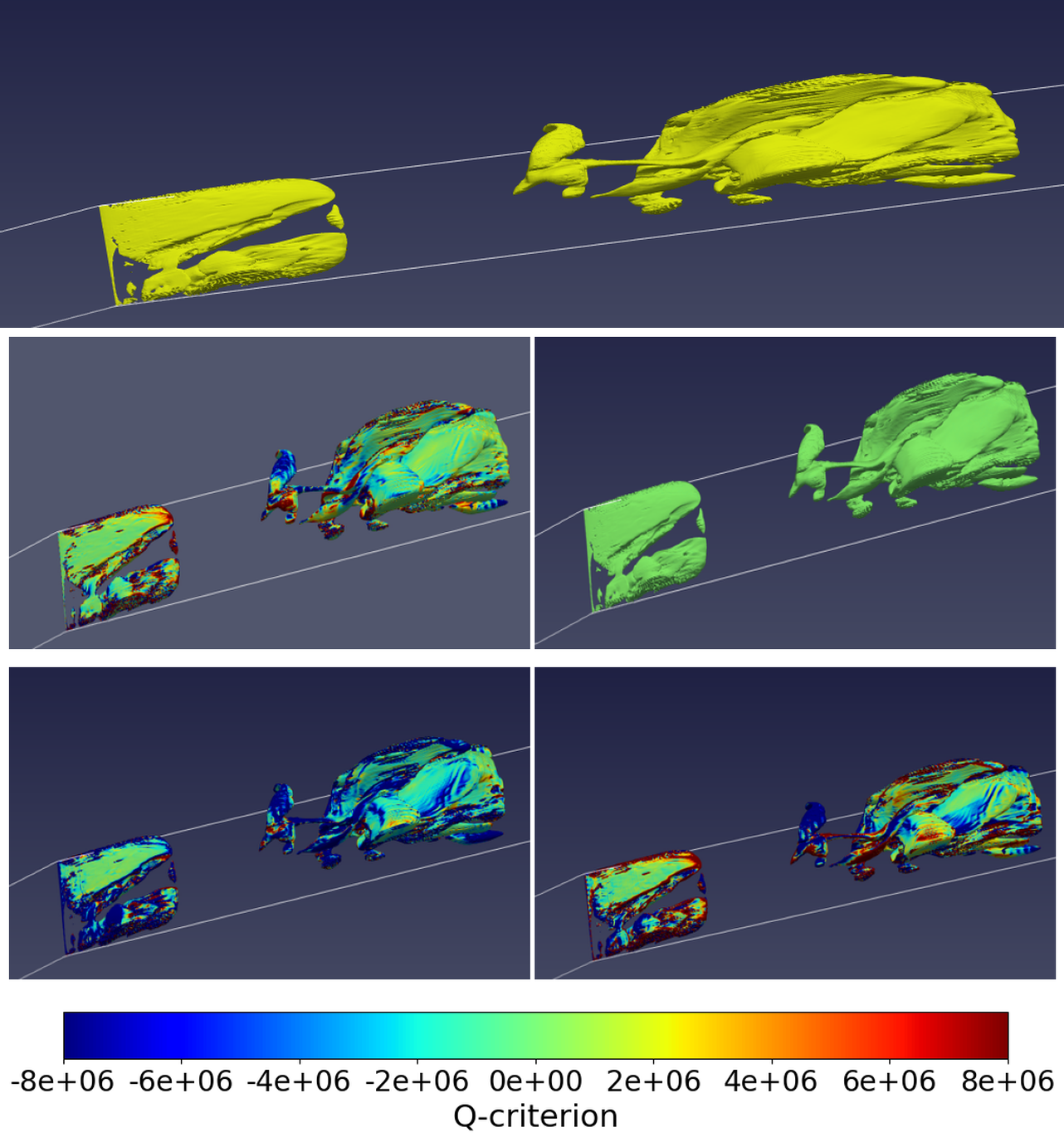}
\caption{Starting top left, in clockwise direction, contours of void fraction at $\alpha=0.6$, vortex stretching, baroclinic torque,vortex dilatation and Q-criterion terms as the incipient cavity starts growing as the detached cavity of the previous cycle flows downstream.}
    \label{fig:ce}
\end{figure*}
Fig \ref{fig:ce} shows the next stage of the cavitation shedding. The previously small incipient cavity grows larger, developing into a cloud shape. Here, the Q-criterion is negative across the primary cavity and the detached cavity of the previous cycle, indicating the vorticity tensor is smaller than the rate of strain tensor. However, it is observed that the vortex stretching and dilatation terms continue to dominate the vorticity tensor compared to the baroclinic torque. 

\begin{figure*}[h]
\centering
\includegraphics[width=\textwidth, height=14cm]{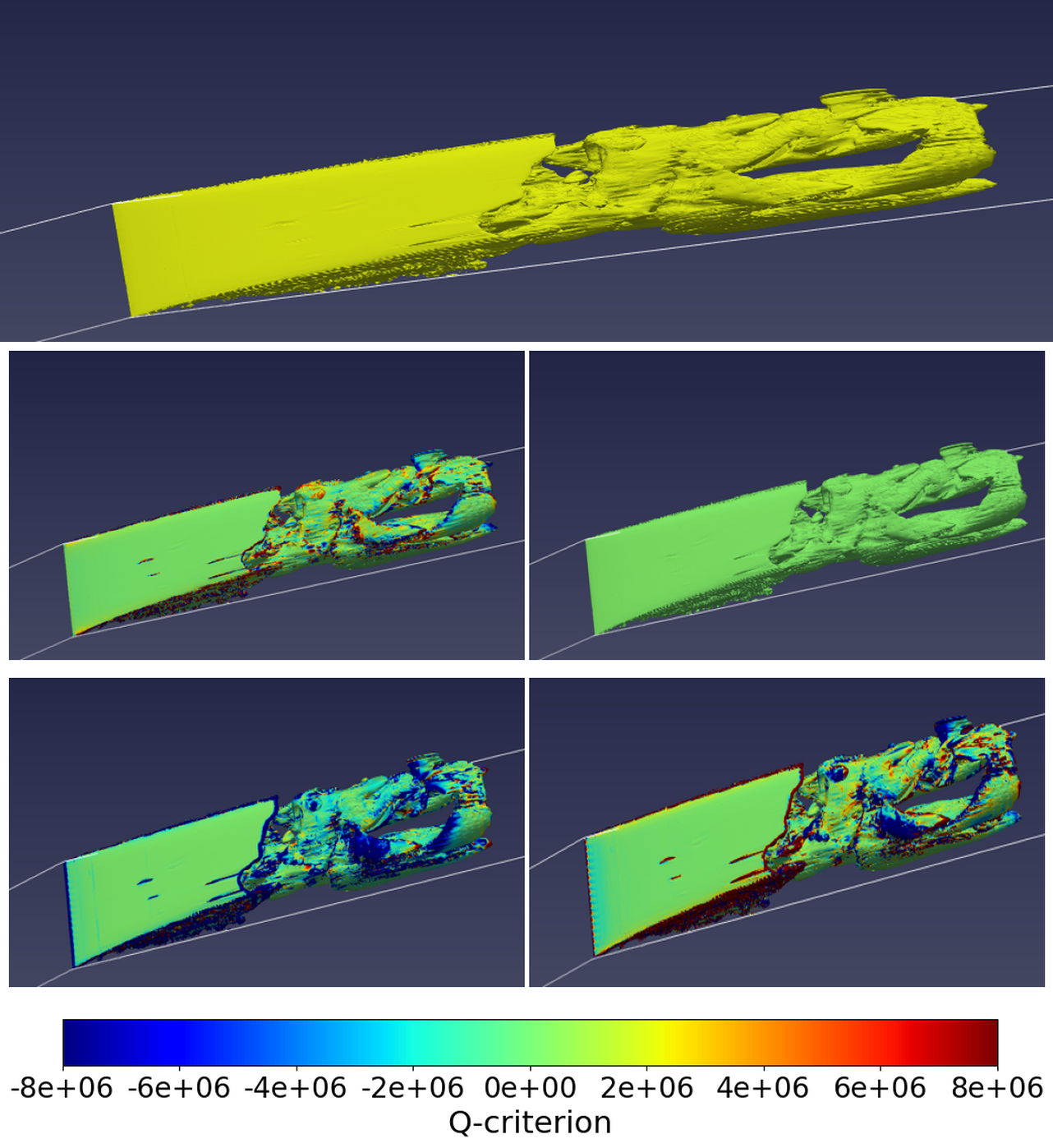}
\caption{Starting top left, in clockwise direction, contours of void fraction at $\alpha=0.6$, vortex stretching, baroclinic torque,vortex dilatation and Q-criterion terms as the primary cavity reaches its maximum size while the secondary cavity of the previous cavity collapses.}
    \label{fig:cm}
\end{figure*}

\begin{figure*}[h]
\centering
\includegraphics[width=\textwidth, height=14cm]{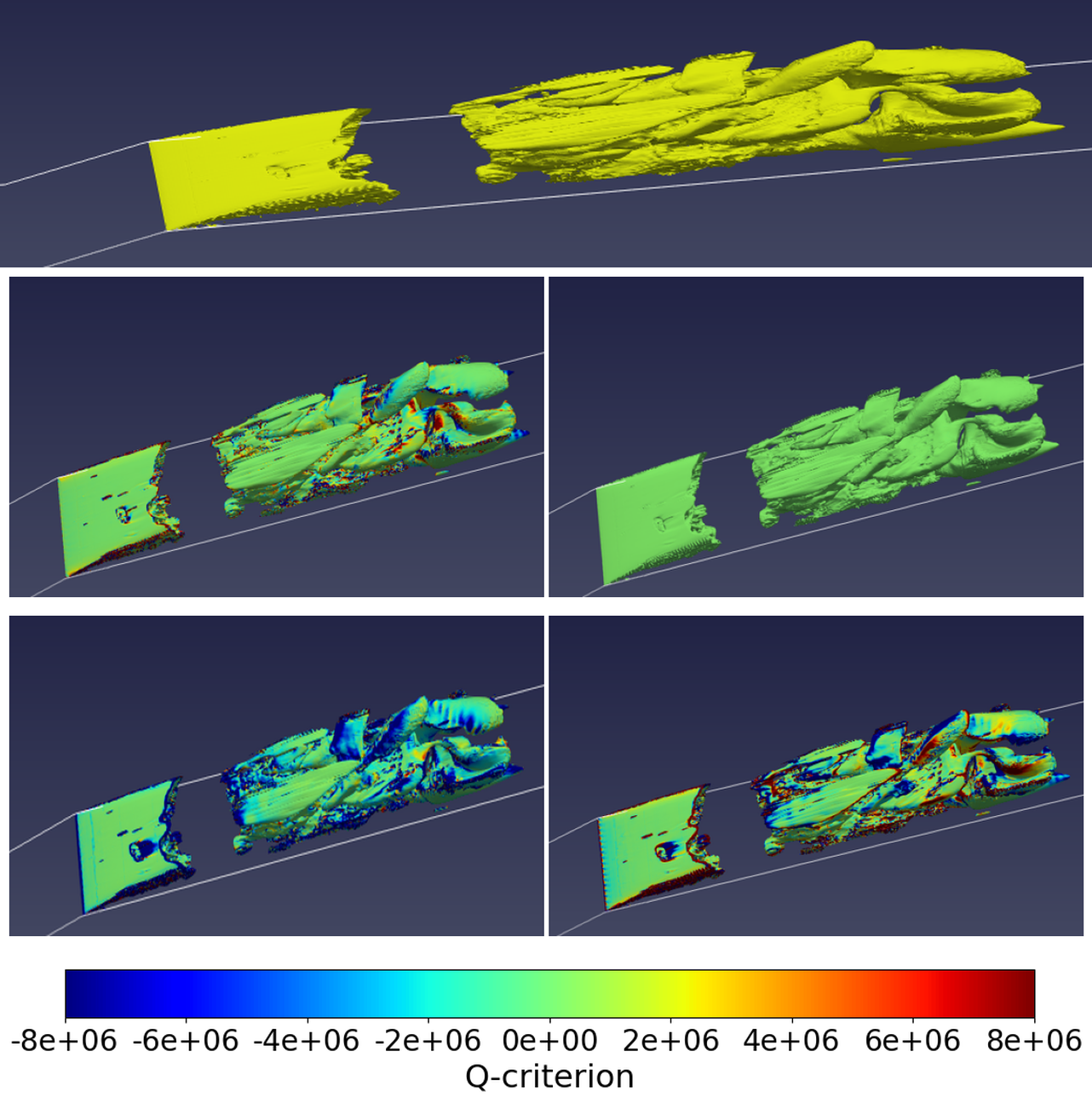}
\caption{Starting top left, in clockwise direction, contours of void fraction at $\alpha=0.6$, vortex stretching, baroclinic torque,vortex dilatation and Q-criterion terms as the primary cavity detaches into a reduced primary cavity at the throat and a detached secondary cavity shed downstream}
    \label{fig:cd}
\end{figure*}

\begin{figure*}[h]
\centering
\includegraphics[width=\textwidth, height=14cm]{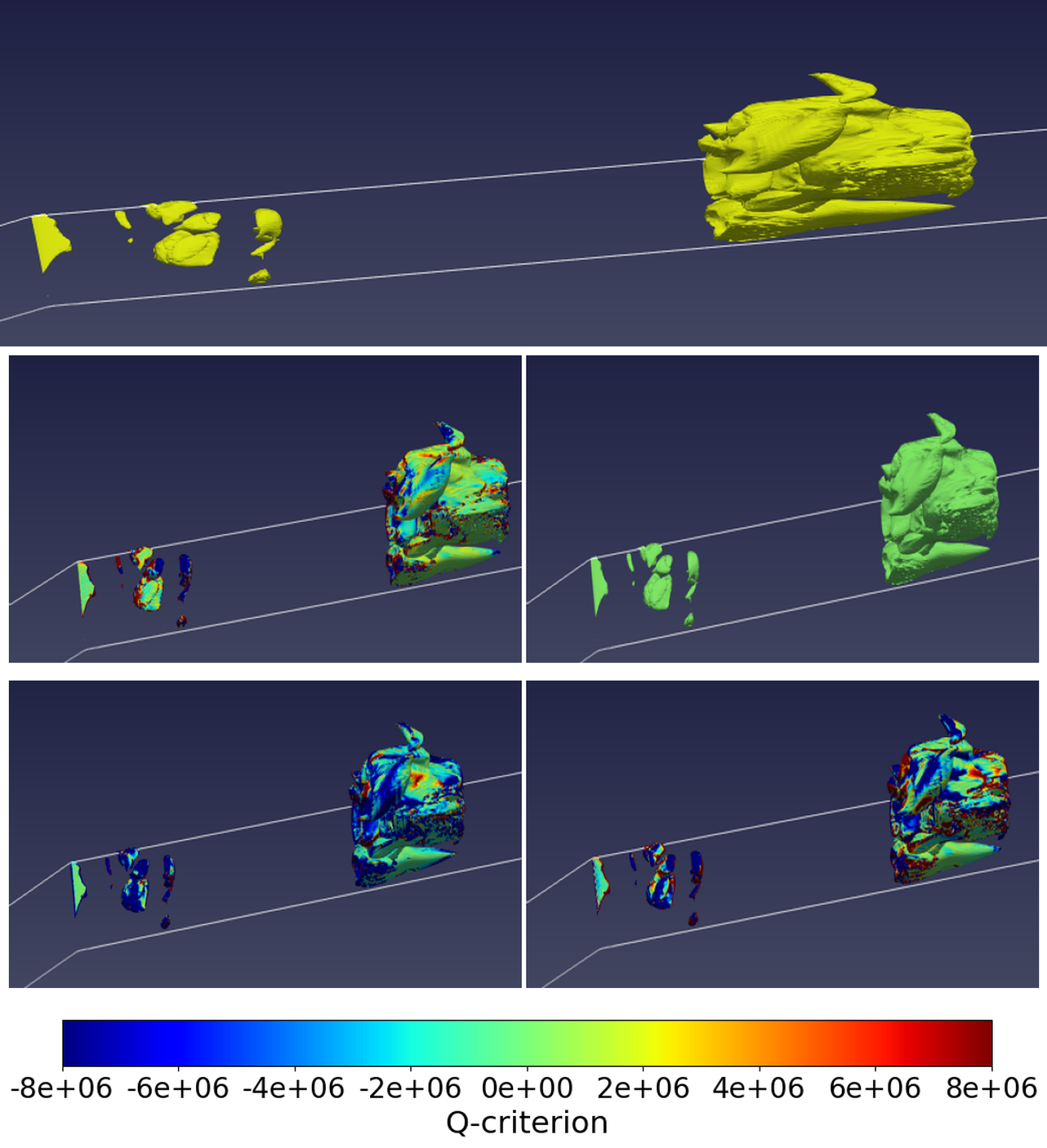}
\caption{Starting top left, in clockwise direction, contours of void fraction at $\alpha=0.6$, vortex stretching, baroclinic torque,vortex dilatation and Q-criterion terms as the shed cavity flow downstream while a new cavity starts forming at the venturi throat.}
    \label{fig:cc}
\end{figure*}
Fig \ref{fig:cm} shows the Q-criterion and the vorticity equation terms as the cavity reaches its maximum size. Here, it is observed that the secondary cavity of the previous cycle has collapsed upon exiting the low-pressure region. The Q-criterion is distributed messily throughout the cavity with the vortex stretching and vortex dilatation term also displaying similar behaviours: the terms are positive along the cavity interface but negative inside the cavity, showing the very drastic change in vorticity from the cavity interface to the center of the cavity. As the cavity reaches its maximum size, it encounters a re-entrant jet travelling in the direction opposite to the main flow direction that breaks the cavity into two, a reduced primary cavity at the throat and a detached secondary cavity downstream (Fig \ref{fig:cd}). Here, the Q-criterion is considerably high near the cavity interface but dominated by the strain rate tensor inside the cavity. While the vortex stretching term remains positive at the cavity-water interface at the point of cavity detachment, the vortex dilatation term gets extremely at the trailing edge of the primary cavity and leading edge of the secondary, rolled-up cavity. It can be observed, the shedding of the primary cavity leads to generation of vortices in both direction. The figures demonstrate the dramatic change of vortices caused by cloud cavitation and the resulting interplay between strain rate and the vorticity tensors and how the expansion and contraction of the fluid element overall dominates the process.  The shed cavity flows downstream and collapses, upon exiting the low pressure, as seen in Fig \ref{fig:cc}. At the same instant, a small incipient cavity is observed at the throat. Thus, the periodic shedding of cavity clouds is observed. With the collapse of the cloud cavity, the Q-criterion is positive in very few areas, demonstrating the role of strain rate tensor in the process. In the areas where vorticity is observed, near the cavity-water interface of the detached cloud cavity, it is observed the vortex shedding dominates the process. While studies \cite{laberteaux2001partial} have postulated that baroclinic torque might be responsible for vorticity production during cloud cavity collapse, baroclinic torque has not been observed as extremely influential in vorticity generation in these numerical simulations. Throughout the stages of cavitation, it is observed that the velocity gradients and the growth $\And$ shrinking of the fluid element resulting due to cavitation influence the vortex formation thus providing insights into the cavitation-vortex interaction.    

\subsection{Cavity dynamics at the local scale}
While the cavity dynamics at global scale is discussed, it is important to investigate if turbulence modelling with AMR is able to capture the cavity dynamics at the local scale.    

\begin{figure}[!htb]
  \centering
\includegraphics[width=1\textwidth]{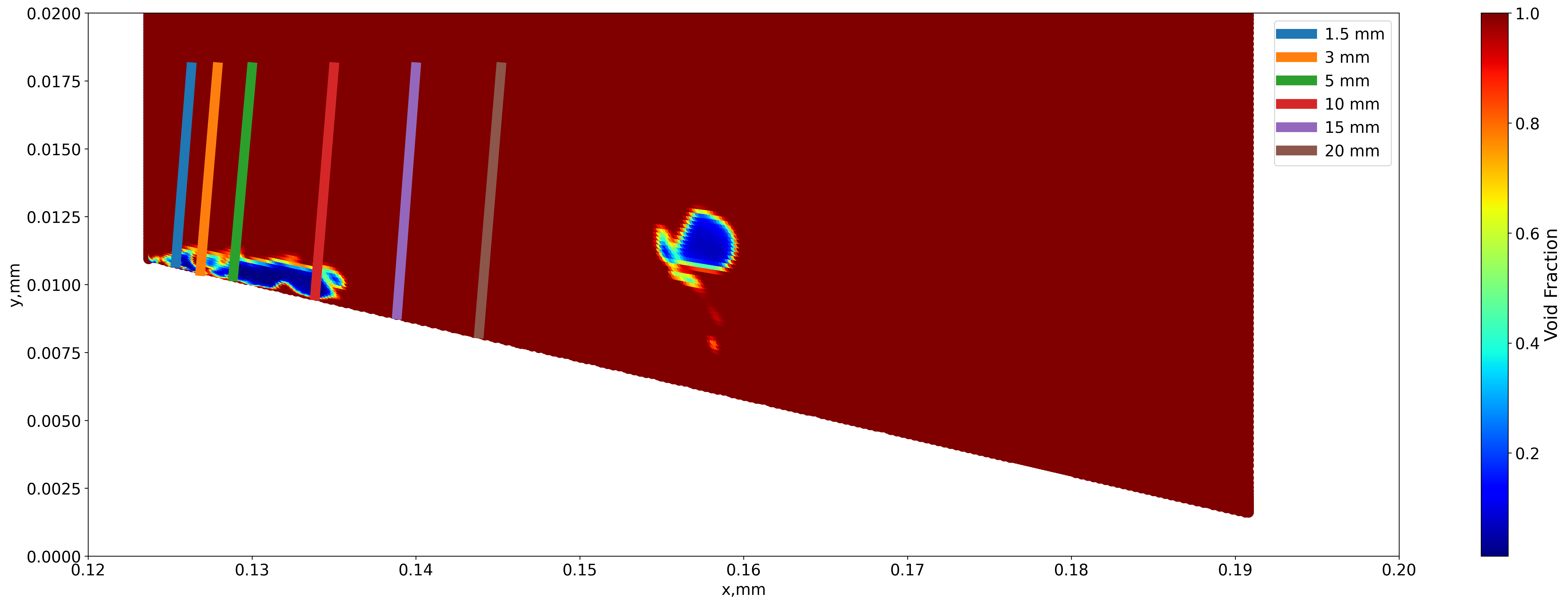}
  \caption{Profile stations in mid-plane of venturi for local analysis. All the stations are within the primary cavity thus capturing significant cavity dynamics. The color plot corresponds to the void fraction: the red areas indicate water while the blue region indicates the vapor cloud cavity }
    \label{fig:profiles}
\end{figure}

\begin{figure}[!htb]
  \centering
\includegraphics[scale=0.27]{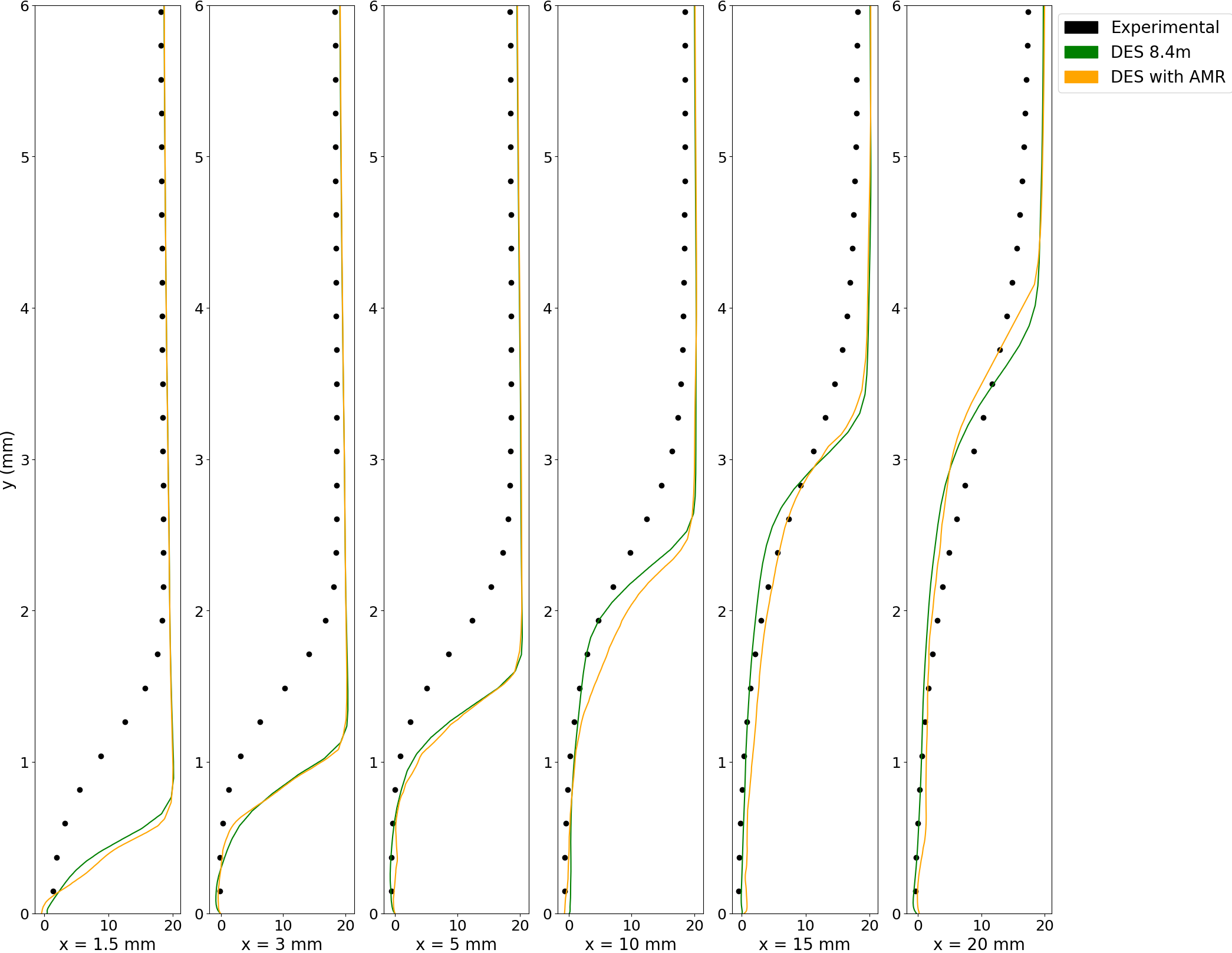}
  \caption{Comparison for velocity profiles in stream-wise direction. Here, the black dots represent the experimental data while the green line represents the non-AMR DES calculation on a 8.4 million cell mesh. }
    \label{fig:vel_streamwise}
\end{figure}

\begin{figure}[h]
  \centering
\includegraphics[scale=0.3]{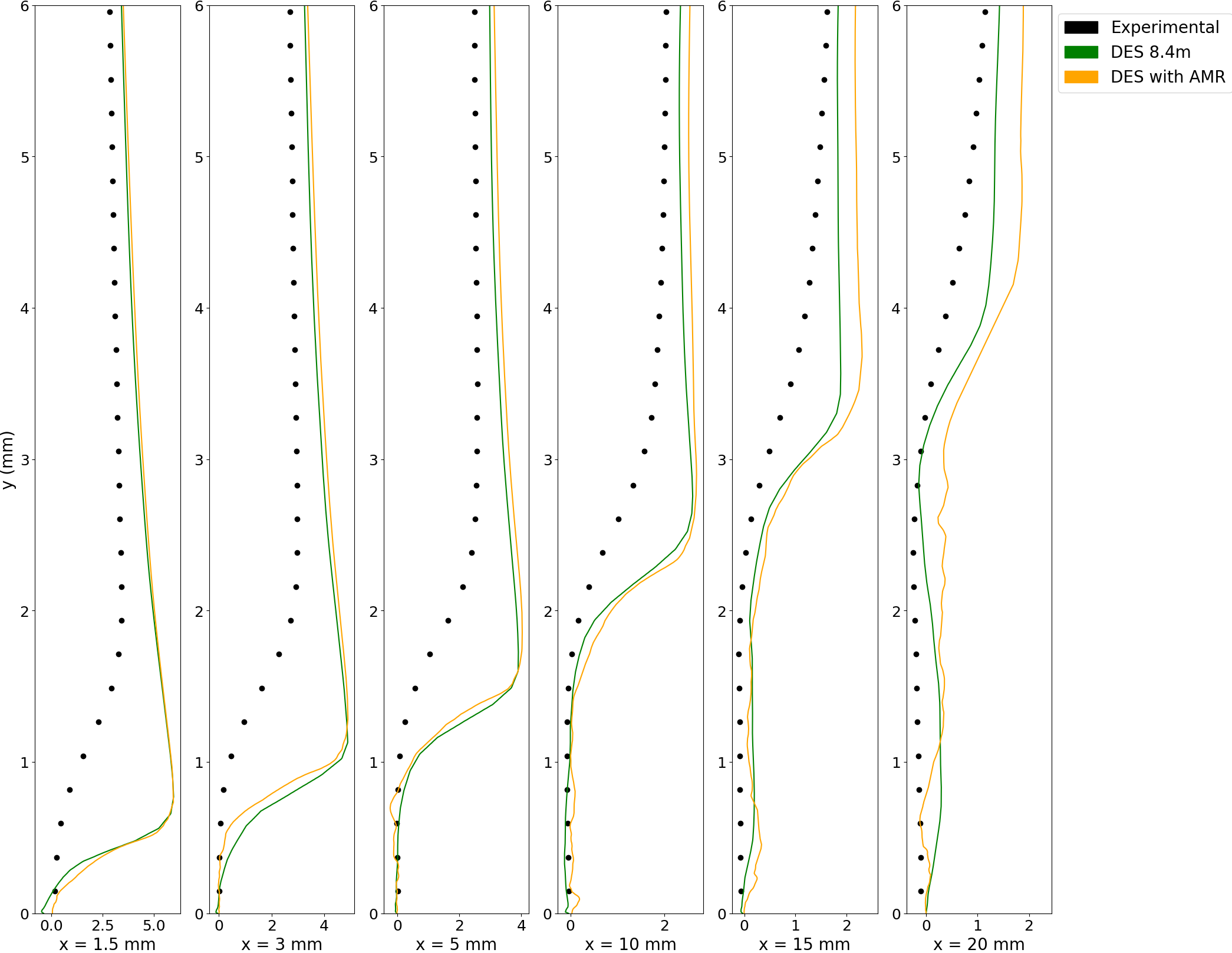}
  \caption{Comparison for velocity profiles in wall direction. Here, the black dots represent the experimental data while the green line represents the non-AMR DES calculation on a 8.4 million cell mesh.}
    \label{fig:vel_wall}
\end{figure}

Fig \ref{fig:profiles} shows the venturi nozzle at cut at mid-plane about the z-axis with the profile stations for local analysis. These stations are 1.5mm, 3 mm, 5mm, 10 mm, 15 mm and 20 mm from the throat respectively. The profile stations have been placed at varying distance from the throat to capture the cavity and investigate if the numerical results are able to simulate the turbulence properties as observed in experiments. Figs \ref{fig:vel_streamwise} and \ref{fig:vel_wall} shows the profiles for time-averaged velocities in both stream-wise and wall directions respectively. The black dots represent the experimental data, taken from Ge \textit{et al.} \cite{ge2021cavitation}. Near the throat, the stream-wise velocity near the venturi's diverging wall is less than the flow velocity but jumps considerably between 1 and 2 mm from the wall. Downstream, the jump is much less pronounced, with the velocity increasing steadily. Both models are able to predict the stream-wise time-averaged velocity well, with slight differences in the 10 mm profile. Extending the discussion to the velocity in wall-direction, different results are obtained. The velocity jump away from the wall is less pronounced throughout the profiles as compared to the velocity in the stream-wise direction. The numerical simulations show considerable discrepancies near the throat as they predict a sharper velocity jump slightly closer to the wall than observed in the experiments. This indicates that while the simulations are able to reproduce the same cavity length as measured in the experiments, they are unable to reproduce the same cavity width at the throat. This discrepancy could stem from the cavitation modelling strategy. In addition, subtle differences appear in the downstream profiles, starting from the 10 mm profile. Away from the wall, the AMR mesh predicts higher velocity than the non-AMR mesh and experimental data. At the farthest profile, the 20 mm one, the AMR profile shows some minor pulsations close to the bottom wall of the venturi which are not captured by the non-AMR mesh. These pulsations could be attributed to the smaller cavity clouds formed after the primary cavity detaches due to the re-entrant jet rushing upwards.    


\begin{figure}[h]
  \centering
\includegraphics[scale=0.3]{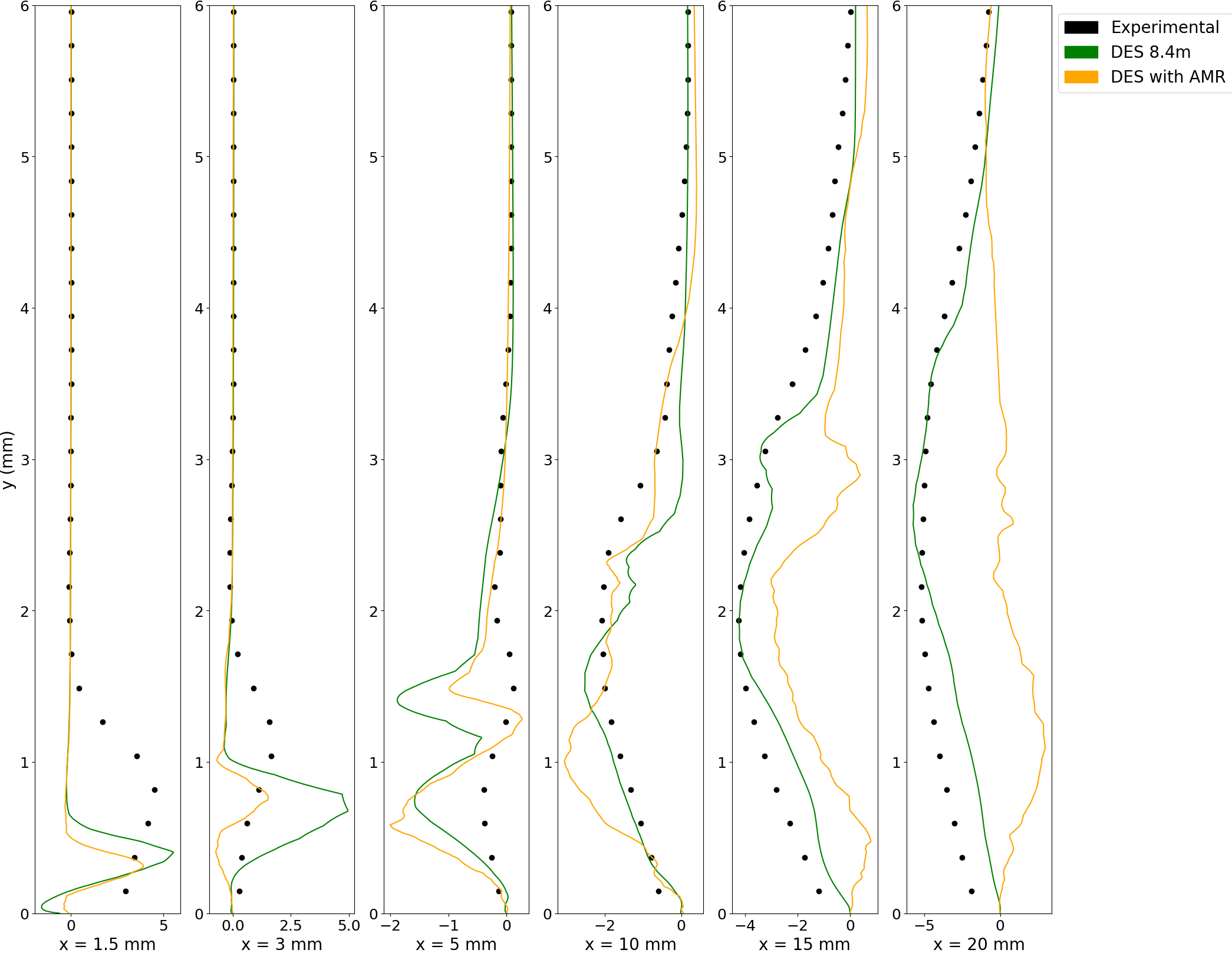}
  \caption{Comparison for Reynolds stress (u'v') profiles. Here, the black dots indicate experimental data while the green line corresponds to the non-AMR DES calculation of 8.4 million cells. The orange line shows the results of the DES coupled with the AMR calculation.  }
    \label{fig:Re}
\end{figure}
\begin{figure}[h]
  \centering
\includegraphics[scale=0.3]{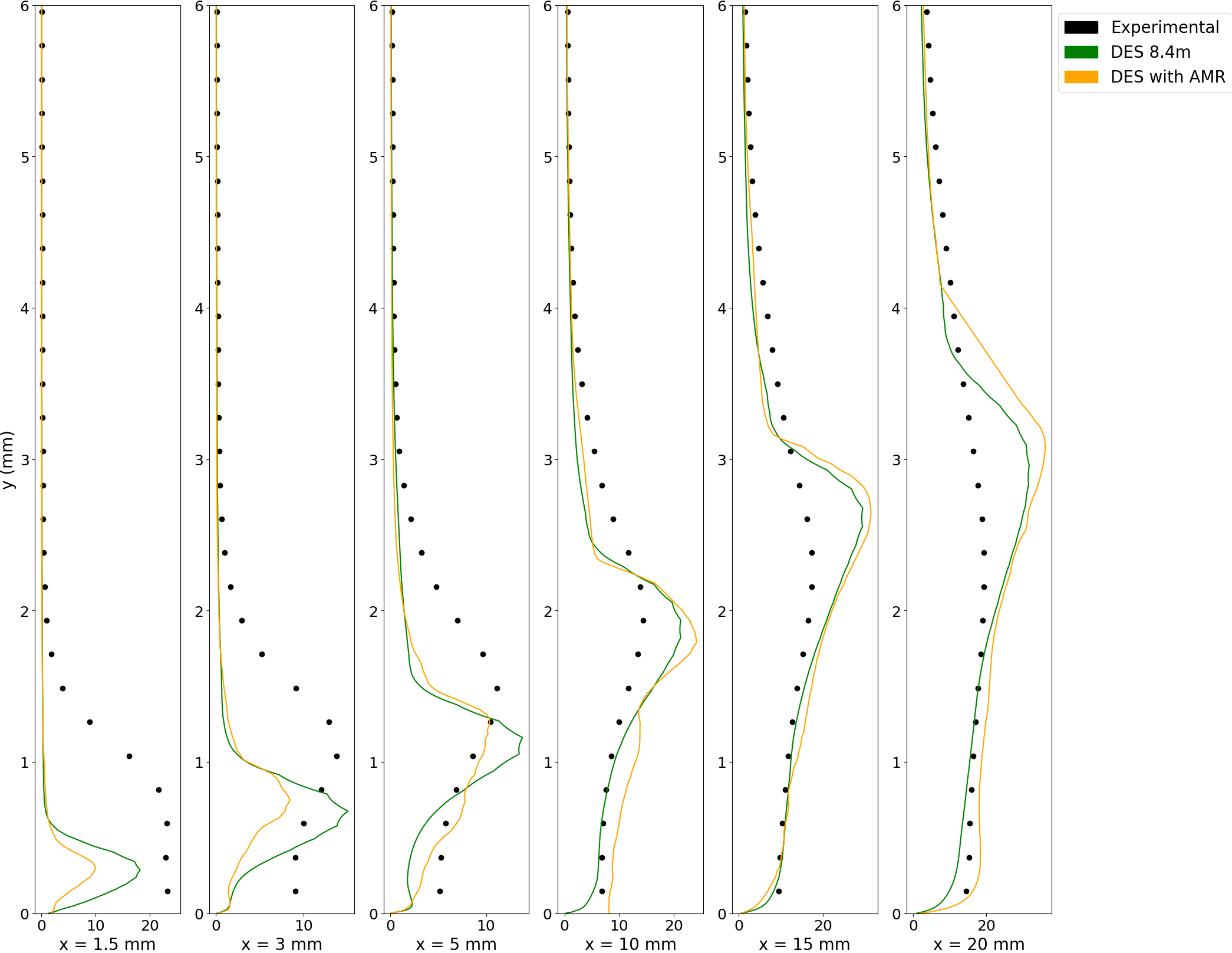}
  \caption{Comparison for Turbulent Kinetic energy (TKE).}
    \label{fig:TKE}
\end{figure}

The analysis is extended to turbulence data like the Reynolds shear stress and the Turbulent Kinetic Energy (TKE) as well, in Figs \ref{fig:Re} and \ref{fig:TKE}. It is observed at the throat, that the AMR calculation predicts the Reynolds stress closer to the experimental data as compared to the non-AMR mesh which over-predicts the Reynolds shear stress. This co-relation extends until the 5 mm profile. It is posited that since the AMR is a function of low void fraction, the grid near the throat becomes considerably more refined than the standard mesh. As a result, the area is modelled by LES rather than URANS, as is the case in DES models, thus leading to better results as compared to standard meshes. However, downstream predictions show considerable discrepancies as the AMR mesh under-predicts the Reynolds stress while the standard mesh performs comparatively better. Downstream, the absence of low void fraction results in a coarser mesh as compared to the region close to the throat. The coarser mesh results in the modelling of the region by URANS rather than LES, thus under-predicting the turbulence stress as previously observed in URANS simulations. The TKE plots show similar results. As we go downstream, the TKE increases continuously and farther away from the bottom wall. The AMR mesh under-predicts the TKE at the profiles near the throat but are able to predict the TKE much better away from the throat. The AMR mesh results are similar to the results displayed by the non-AMR mesh that contained 8.4 million cells throughout the calculation. Thus, while the AMR is reproducing the same cavity dynamics, it provides insights into cavitating flow with considerably much less computational time than standard refined mesh, at the local scale scale as well.  

\section{Conclusions}

This paper employs the Adaptive Mesh Refinement (AMR), coupled with DES model to simulate the unsteady cavitating flow in a converging-diverging (venturi) nozzle. The Merkle model is used to model for cavitation modelling. The vapor cavity structures, the mechanisms of cloud shedding and cavitation-vortex interaction are examined alongside the cavitation-turbulence coupling on the local level. The main conclusions are as follows:
\begin{enumerate}
    \item The numerical simulations show that cloud cavitation is a periodic process where a cavity initiates at the throat, followed by its growth until it reaches a maximum size before being pinched-off by a re-entrant jet. Further analysis of the cavitation-vortex interaction using the terms comprising vorticity transport equation demonstrated the extensive domination of the vortex stretching and the vortex dilatation terms throughout the cavitation cycle. This indicates the  strong influence of the velocity gradients and the growth of the fluid element in the vorticity and therefore, the strong interplay between cavitation and vortex formation. The baroclinic torque is much smaller and less influential than other terms. 
    \item Analysis of time-averaged velocities and turbulence-related aspects like, the Reynolds stress and TKE on local profile stations showed the influence of AMR as compared to a non-AMR mesh calculation. While the AMR mesh is able to predict well the velocity in stream-wise direction, it displays some differences in the wall direction velocity in the downstream profiles. However, while the numerical simulations are able to reproduce the cavity lengths as measured in experiments, there is significant difference in the cavity width at the throat, where the numerical simulations predict a thinner cavity. 
    \item The AMR mesh calculation is able to predict the Reynolds stress data near the throat much better than the standard mesh calculation. However, the prediction worsens downstream, where the AMR calculation under-predicts the Reynolds stress. This is a consequence of the absence of low void fraction downstream, resulting in lower mesh refinement and the region being subsequently modelled by URANS. The TKE plots show both calculations display similar results and reproducing TKE downstream in a close agreement with experiments, thus highlighting the use of AMR as a tool to speed up calculations for unsteady cavitating flows. 
    \item Future avenues include running the AMR simulation with more refinement levels to investigate the results in a better way at a reduced computational cost. However, as this study sheds more light on turbulence-cavitation interaction, future AMR calculations should focus on grid refinement based on turbulence data rather than void fraction in order to model the turbulence-cavitation interplay more accurately, throughout the domain.   

\end{enumerate}

\section{Acknowledgements}
This work was supported by the Office of Naval Research, USA [grant number N00014-18-S-B001], the Macao Young Scholars Program (Project code: AM2022003) and Priority Postdoctoral Projects in Zhejiang Province (Project Code:341781). The authors would like to thank the ONR proposal manager Dr. Ki-Han Kim for his support. The authors would like to thank the anonymous reviewers for their feedback to enhance the quality of the manuscript. 

\section{Compliance with ethical standards}
\begin{itemize}
    \item \textbf{Conflict of interest}: The authors declare they have no conflict of interest.
    \item \textbf{Ethical approval}: The article does not contain any studies with human participants or animals performed by any of the authors.
    \item \textbf{Informed consent}: Not applicable 
    \item \textbf{Funding}:This work was supported by the Office of Naval Research, USA [grant number N00014-18-S-B001]. The authors would like to thank the ONR proposal manager Dr. Ki-Han Kim for his support. 
\end{itemize}


\bibliography{sn-bibliography}

\end{document}